\documentstyle[12pt]{article}
\title{\bf 4+1 dimensional homogeneous anisotropic string cosmological models }
\author{B. Mojaveri{\thanks{e-mail:
bmojaveri@azaruniv.edu}}\,\,\,and A. Rezaei-Aghdam{\thanks{e-mail:
rezaei-a@azaruniv.edu}} \\
{\small Department of Physics, Azarbaijan university of Tarbiat
Moallem, 53714-161, Tabriz, Iran .}}
\begin{document}
\maketitle
\begin{abstract}
We present exact solutions of string cosmological models
characterized by five dimensional metrics (with four dimensional
real Lie groups as isometry groups), space independent dilaton and
vanishing torsion. As an example we consider $VII_0\oplus R$ model
and show that it is equivalent to the $4+1$ dimensional cosmological
model coupled to perfect fluid with negative deceleration parameters
(accelerating universe).
\end{abstract}

\newpage
\section{\bf Introduction}

Some of the important problems of primordial cosmology such as, why
is our universe four-dimensional? why does a small vacuum
energy-density seem to survive until today after inflation? and
other problems should find a satisfactory solution in the context of
string cosmology \cite{Brandenberg}. The dynamics of the very early
universe below the string scale may have been determined by the
string effective action \cite{Frad}. The idea that string theory,
with its fundamental length scale, could resolve the big bang
singularity by effective limiting space-time curvature has led,
during the last decade, to a variety of models \cite{Gasperini3}
where the big bang represent a turning (rather than an end)-point in
the history of the early Universe ( see also \cite{Bozza,
Gasperini5}). The general, spatially isotropic and homogeneous, FRW
string cosmology is known and well understood \cite{Copeland}.
Furthermore, it is generally accepted that spatial anisotropy might
have been important in the very early universe and the study of the
string cosmology models that relax the FRW assumptions is therefore
well motivated. The spatially homogeneous anisotropic Bianchi
cosmologies \cite{Barrow, Ba, Gas} have been studied previously and
some inhomogeneous generalizations of these models were presented
\cite{Barrow1}. Bianchi cosmologies admit a three-dimensional real
Lie group of isometries that acts simply-transitively on
three-dimensional, space like orbits \cite{Ellis}. Some
generalization of these models to models with four-dimensional real
Lie groups was also performed \cite{Hervik}.

Here we present exact solutions of string cosmological models
characterized by five dimensional metric (with four dimensional real
Lie group as isometry groups), a dilaton field and vanishing
torsion.
\newline The paper is organized as follows. In section two,
we review the homogenous anisotropic string cosmology in d-dimension
\cite{Gas} and then obtain main equations in $4+1$ dimension. After
then in section three, we obtain metrics over real four dimensional
Lie groups. Subsequently, in section four we solve and obtain exact
solutions of equations of motions for $4+1$ dimensional homogenous
anisotropic string cosmological models related to real four
dimensional Lie algebras. Finally in section five we consider
example $VII_0\oplus R$ and show that it is equivalent with $4+1$
dimensional cosmological coupled to perfect fluid with negative
decelerating parameters (accelerating universe). Some concluding
remarks including $T$- duality discussions are given in section six.
\section{\bf Review of homogeneous anisotropic string cosmology}
We start with string effective action \cite{Frad}:
\begin{equation}
S_{eff}=\int d^{d+1}x \sqrt{-g}e^{\phi}(R-\frac{1}{12}
H_{\mu\nu\rho}H^{\mu\nu\rho}+\partial_{\mu}\phi\partial^{\mu}\phi-\Lambda)
\label{b4'}.
\end{equation}
The equations of motion for this action have the following forms
\begin{eqnarray}
R_{\mu\nu}-\frac{1}{4}H_{\mu\nu}^2-\nabla_\mu\nabla_\nu\phi&=&0,
\label{b1}\\
\nabla^2(e^\phi H_{\mu\nu\lambda})&=&0, \label{b2} \\
-R+\frac{1}{12}H^2+2\nabla^2
\phi+(\partial_\mu\phi)^2+\Lambda&=&0,\label{b3}
\end{eqnarray}
where the contractions $H_{\mu\nu}^2=H_{\mu\kappa\lambda}
{H_{\nu}}^{\kappa\lambda} \, ,
H^2=H_{\mu\nu\lambda}H^{\mu\nu\lambda}$ involve the totally
antisymmetric field strength $H_{\mu\nu\lambda}$, defined in terms
of the potential $B_{\mu\nu}$ as
\begin{eqnarray}
H_{\mu\nu\rho}=\partial_\mu B_{\nu\rho}+\partial_\rho
B_{\mu\nu}+\partial_{\nu} B_{\rho\mu} \label{H}.
\end{eqnarray}
Indeed these equations are the Einstein field equations coupled with
matters $\phi$ and $H$ ( see for instance \cite{Ba}). Note that for
the study of 4+1 dimensional homogeneous string cosmological models,
we first consider it in $d+1$-dimension . We know that with the
assumption of spatial homogeneity, the $d$-dimensional spatial
submanifold is invariant under the action of a d-parametric isometry
group $G$, and the metric of the space-time $G_{\mu\nu}$ can be
factorized in a synchronous frame as \cite{La}:
\begin{eqnarray}
ds^2=G_{\mu\nu}dx^{\mu}dx^{\nu}=-dt^2+
{e_\alpha}^i(x)g_{ij}(t){e_\beta}^j(x)dx^\alpha dx^\beta,
\end{eqnarray}
where $\alpha , \beta=1,...,d$ are world indices in the spatial
submanifold and $i,j$ are indices for the bases of the isometry
group G. Notice that vielbeins ${e_\alpha}^i(x)$ are dependent only
on spatial coordinates \cite{La}. Now, if we consider additional
assumption that the dilaton field be space independent and the
torsion field $H_{\mu\nu\rho}$ vanish then the equations(2)-(4) can
be reduce to ordinary time differential equations for the variables
$g_{ij}(t)$ and $\phi(t)$ \cite{Gas}. Furthermore if we restrict our
analysis to an anisotropic but diagonal matrix form for the
invariant metric $g_{ij}(t)$ \footnote{Note that one can also use
other form of $g_{ij}$; for example for four dimensional Lie groups
one can use results of Ref \cite{Hervik1}.},
\begin{eqnarray}
g_{ij}(t)=a_i^2(t) \delta_{ij},
\end{eqnarray}
where $a_i(t)$ are scale factors; then the (0,0) and (i,i)
components of the equation (2) have the following form respectively
\cite{Gas}:
\begin{eqnarray}
{\sum_{i=1}}^d(\dot{H_i}+ H_i^2)+\ddot{\phi}=0,
\end{eqnarray}
\begin{eqnarray}
\dot{H_i}+H_i{\sum_{k=1}}^d H_k + H_i \dot{\phi}+V_i =0,
\end{eqnarray}
where the dot stands for derivative with respect to $t$; furthermore
(i,0) components give the constraints\footnote{Here there is no
summation over $i$.}
\begin{eqnarray}
{\sum_{k=1}}^d C_{ki}^k( H_i - H_k)=0.
\end{eqnarray}
On the other hand, $i\neq j$ components of equation (2) give a set
of another constraint equations
\begin{eqnarray}
R_{i,j}=0,\qquad i\neq j,
\end{eqnarray}
and the dilaton equation (4) has the following form:
\begin{eqnarray}
-2\ddot{\phi}-\dot{\phi}^2-2\dot{\phi}{\sum_{k=1}}^d H_k
-{\sum_{k=1}}^d V_k- ({\sum_{k=1}}^d H_k)^2
-{\sum_{k=1}}^d{H_k}^2-2{\sum_{k=1}}^d \dot{H_k}=0.
\end{eqnarray}
Note that in the above equations $H_i=\frac{\dot{a_i}}{a_i}$ are
Hubble coefficients, $C_{ij}^k$ are structure constant of the
isometry Lie algebras $\textsf{g}$ and $V_k(a_i)$ are potential
functions that depend on the type of Lie group. In this way instead
of equations $(2-4)$ one can use equations $(8-12)$ and solve them
to obtain homogeneous string cosmological models over space-times
with isometry group $G$.\newline

Now, for $4+1$ dimensional models the isometry groups are four
dimensional real Lie groups. There are various classifications for
these Lie algebras \cite{Mu, Pet, Pat, Mac}. Here in this paper we
use the Patera et \cite{Pat} classification in the next section.
Now, for the four dimensional spatial submanifold one can
introduce mean radius $a$ in term of scale factors $a_i$ as
follows:
\begin{eqnarray}
a^4=a_1a_2a_3a_4,
\end{eqnarray}
then using equations (8),(9),(12) we have
\begin{eqnarray}
\ddot{\phi}+\dot{\phi}^2+4\dot{\phi}\frac{\dot{a}}{a}=0.
\end{eqnarray}
On the other hand, introducing new time coordinate $\tau$ \cite{Ba}
\begin{eqnarray}
d\tau=a^{-4} e^{-\phi}dt,
\end{eqnarray}
the equation (14) can be rewritten as
\begin{eqnarray}
\phi''=0,
\end{eqnarray}
where prime stands for derivative with respect to $\tau$. The
general solution of the above equation is
\begin{eqnarray}
\phi=N \tau,\quad or\quad e^{-\phi}=e^{-N \tau},
\end{eqnarray}
where $N$ is a constant. Furthermore, with this new time coordinate
the equation (9) may be rewritten as
\begin{eqnarray}
{\ln(a_i^2 e^{\phi})}''+2a^8 e^{2\phi}V_i =0.
\end{eqnarray}
On the other hand, equation (12) can be reexpressed as the following
initial value equations:
\begin{eqnarray}
\sum_{i<j}{\ln(a_i^2 e^{\phi})}'{\ln(a_j^2 e^{\phi})}'+2\sum_iV_i
a^8 e^{2\phi}-2\sum_i\frac{{a_i}'}{a_i}\phi'-4{\phi'}^2=0.
\end{eqnarray}
Now, it is enough to solve equation (18) with the constraints (10)
and (11)and initial values equations (19). In order to solve
equations (18) we must obtain potentials $V_i(a_j)$.  In the next
section we obtain metrics over four dimensional Lie groups, then in
section four we obtain these potentials and solve equations (18).

\section{\bf Four dimensional real Lie algebras and metrics over their Lie groups}
$\; \; \; \;$There are several classifications for real four
dimensional Lie algebras (for instance see \cite{Mu}, \cite{Pet},
\cite{Pat}, \cite{Mac}). Because in most cosmological applications
the Patera and et al classification has been used (see for
instance \cite{Hervik, Hervik1} and references), here we apply
this classification (Table 1). Now, to obtain the metrics of 4+1
dimensional anisotropic homogenouse space-times according to (6),
we must first calculate vielbeins ${e_\alpha}^i(x)$ for Lie groups
of table 1. To this end, we use the following relation:
\begin{eqnarray}
g^{-1}dg ={e_\alpha}^i(x)X_i d x^\alpha, \quad\quad g\in G.
\end{eqnarray}
With the following parameterizations for the real four dimensional
Lie groups $G$:
\begin{eqnarray}
g = e^{x_1 X_1} e^{x_2 X_2}e^{x_3 X_3}e^{x_4 X_4},
\end{eqnarray}
where $\{X_i\}$ and $\{x_i\}$ are generators and coordinates of Lie
group, respectively. Then in general, for left invariant Lie algebra
valued one forms we have:
\begin{eqnarray}
&&\hspace{-5mm}g^{-1}dg=dx_1 e^{-x_4 X_4}e^{-x_3 X_3}(e^{-x_2
X_2}X_1 e^{x_2 X_2}) e^{x_3 X_3}e^{x_4 X_4} + dx_2 e^{-x_4
X_4}e^{-x_3 X_3}X_2 e^{x_3
X_3}e^{x_4 X_4}\nonumber\\
&&\hspace{10mm}+dx_3 e^{-x_4 X_4}X_3e^{x_4 X_4}+dx_4X_4.
\end{eqnarray}
Now, we need to calculate expressions such as $e^{-x_i X_i}X_j
e^{x_i X_i}$. Indeed, in\cite{Jafarizadeh} we have shown that:
\begin{eqnarray}
e^{-x_i X_i}X_j e^{x_i X_i} = (e^{x_i{\cal X}_i})_j^{\;\;k} X_k,
\end{eqnarray}
where summation over index k is assumed and there is no summation
over index i. In this formula $({\cal X}_i)_l^{\; \;j} =
-{f_{il}}^j$ are the adjoint representation of the Lie algebra. In
this way one can calculate all left invariant one forms, which are
shown in Table 1.\\\\

\newpage

\begin{center}
\hspace{0.01mm}{{\bf Table 1}}:\hspace{1mm}4-dimensional real Lie
algebras, their isometry Lie algebras and left invariant one forms.
\footnote{Note that for Lie algebras written as direct sum of three
dimensional Lie algebras with one dimensional Lie algebras we write
their isomorphism with Bianchi $\oplus R$ Lie
algebras.}\\ \hspace{-1mm}\\
    \begin{tabular}{l l l l  l l p{15mm} }
    \hline\hline
   \vspace{-1mm}
{\scriptsize ${\bf Algebra}$ }& {\scriptsize Non-zero commutation
relations }&{\scriptsize Isometry }&{\scriptsize $g^{-1} dg$}
\\\hline

\vspace{1mm}

{\scriptsize ${ 4A_1}$}& {\scriptsize $[X_i,X_j]=0$}&
{\scriptsize $I\oplus R$} & {\scriptsize $dx^i X_i$}\\

\vspace{-1mm}

{\scriptsize ${ A_2+2A_1}$}& {\scriptsize$[X_1,X_2]=-X_2-X_3,
[X_3,X_1]=X_2+X_3$}&
{\scriptsize $X_1,X_4$} & {\scriptsize $dx_1[X_1-(x_2+x_3)(X_2+X_3)]+dx_2X_2+dx_3(X_3$}\\

\vspace{1mm}

& &  & {\scriptsize $+x_4 X_4)+dx_4X_4$}\\

\vspace{1mm}

{\scriptsize ${ 2A_2}$}& {\scriptsize$[X_1,X_2]=X_2,[X_3,X_4]=X_4$}&
{\scriptsize $X_1,X_3$} &{\scriptsize$dx_1(X_1+x_2X_2)+dx_2X_2+dx_3(X_3+x_4X_4)+dx_4X_4$}\\

\vspace{1mm}

& &  & {\scriptsize $$}\\

\vspace{1mm}

{\scriptsize ${II\oplus R}$}& {\scriptsize$[X_2,X_3]=X_1$}&
{\scriptsize $X_1,X_2,X_4$} &
{\scriptsize $dx_1X_1+dx_2(X_2+x_3X_1)+dx_3X_3+dx_4X_4$} \\

\vspace{-1mm}

{\scriptsize ${IV\oplus R}$}&
{\scriptsize$[X_1,X_2]=-X_2+X_3,[X_1,X_3]=-X_3$}&{\scriptsize $X_1,X_4$} & {\scriptsize $dx_1 [X_1-x_2X_2+(x_2-x_3)X_3]+dx_2X_2+dx_3X_3$}\\

\vspace{1mm}

& &  & {\scriptsize $+dx_4X_4$}\\

\vspace{1mm}

{\scriptsize ${V\oplus R}$}&
{\scriptsize$[X_1,X_2]=-X_2,[X_1,X_3]=-X_3$}& {\scriptsize
$X_1,X_4$} &
{\scriptsize $dx_1\left(X_1-x_2X_2-x_3X_3\right)+dx_2X_2+dx_3X_3+dx_4X_4$}\\

\vspace{-1mm}

{\scriptsize ${VI_0\oplus R}$}&
{\scriptsize$[X_1,X_3]=X_2,[X_2,X_3]=X_1$}& {\scriptsize
$X_1,X_2,X_4$} &
{\scriptsize $dx_1(X_1\cosh{x_3}+X_2\sinh{x_3})+dx_2(X_2\cosh{x_3}$}\\

\vspace{1mm}

& &  & {\scriptsize$+X_1\sinh{x_3})+dx_3X_3+dx_4X_4$}\\

\vspace{-1mm}

{\scriptsize ${VI_a\oplus R}$}&
{\scriptsize$[X_1,X_2]=-aX_2-X_3,[X_3,X_1]=X_2+aX_3$}& {\scriptsize
$X_1,X_4$} &
{\scriptsize $dx_1[X_1-(ax_2+x_3)X_2-(ax_3+x_2)X_3]+dx_2X_2$}\\

\vspace{1mm}

& &  & {\scriptsize $+dx_3X_3+dx_4X_4$}\\

\vspace{-1mm}

{\scriptsize ${VII_0\oplus R}$}&
{\scriptsize$[X_1,X_3]=-X_2,[X_2,X_3]=X_1$}& {\scriptsize
$X_1,X_2,X_4$} &
{\scriptsize $dx_1(X_1\cos{x_3}-X_2\sin{x_3})+dx_2(X_2\cos{x_3}+X_1\sin{x_3})$}\\

\vspace{1mm}

& &  & {\scriptsize$+dx_3X_3+dx_4X_4$}\\

\vspace{-1mm}

{\scriptsize ${VII_a\oplus R}$}&
{\scriptsize$[X_1,X_2]=-aX_2+X_3,[X_3,X_1]=X_2+aX_3$}& {\scriptsize
$X_1,X_4$} &
{\scriptsize $dx_1[X_1-(ax_2+x_3)X_2-(-ax_3+x_2)X_3]+dx_2X_2$}\\

\vspace{1mm}

& &  & {\scriptsize $+dx_3X_3+dx_4X_4$}\\

\vspace{-1mm}

{\scriptsize ${VIII\oplus R}$}&
{\scriptsize$[X_1,X_3]=-X_2,[X_2,X_3]=X_1,[X_1,X_2]=-X_3$}&
{\scriptsize $X_1,X_4$} &
{\scriptsize $dx_1(X_1\cosh{x_2}\cos{x_3}-X_2\cosh{x_2}\sin{x_3}-X_3\sinh{x_2})$}\\

\vspace{1mm}

& &  &{\scriptsize$+dx_2(X_2\cos{x_3}+X_1\sin{x_3})+dx_3X_3+dx_4X_4$}\\

\vspace{-1mm}

{\scriptsize ${IX\oplus R}$}&
{\scriptsize$[X_1,X_3]=-X_2,[X_2,X_3]=X_1,[X_1,X_2]=X_3$}&
{\scriptsize $X_1,X_4$} &
{\scriptsize $dx_1(X_1\cos{x_2}\cos{x_3}-X_2\cos{x_2}\sin{x_3}+X_3\sin{x_2})$}\\

\vspace{1mm}

& &  &{\scriptsize$+dx_2(X_2\cos{x_3}+X_1\sin{x_3})+dx_3X_3+dx_4X_4$}\\

\vspace{-1mm}

{\scriptsize${A_{4,1}}$}&{\scriptsize$[X_2,X_4]=X_1,[X_3,X_4]=X_2$}&
{\scriptsize$X_1,X_2,X_3$} &{\scriptsize $dx_1X_1+dx_2(X_1x_4+X_2)+dx_3(X_3+\frac{x_4^2}{2}X_1+X_2x_4)$}\\

\vspace{1mm}

& &  &{\scriptsize$+dx_4X_4$}\\

\vspace{1mm}

{\scriptsize${{A^a}_{4,2}}$}&{\scriptsize$[X_1,X_4]=aX_1,[X_2,X_4]=X_2,[X_3,X_4]=X_2+X_3$}&
{\scriptsize $X_1,X_2,X_3$} &
{\scriptsize $dx_1X_1e^{ax_4}+dx_2X_2e^{x_4}+dx_3(X_3+X_2)e^{x_4}+dx_4X_4$}\\

\vspace{-1mm}

{\scriptsize ${{A^1}_{4,2}}$}&
{\scriptsize$[X_1,X_4]=aX_1,[X_1,X_4]=X_1
,[X_2,X_4]=X_2,$}&{\scriptsize $X_1,X_2,X_3$} &
{\scriptsize $dx_1X_1e^{x_4}+dx_2X_2e^{x_4}+dx_3(X_3+X_2)e^{x_4}+dx_4X_4$}\\

\vspace{1mm}

{\scriptsize
$$}&{\scriptsize$[X_3,X_4]=X_2+X_3$}&{\scriptsize$$}&{\scriptsize$$}\\

\vspace{1mm}

{\scriptsize ${A_{4,3}}$}&
{\scriptsize$[X_1,X_4]=X_1,[X_3,X_4]=X_2$}&{\scriptsize$X_1,X_2,X_3$}&
{\scriptsize $dx_1X_1e^{x_4}+dx_2X_2+dx_3(X_3+x_4X_2)+dx_4X_4$}\\

\vspace{-1mm}

{\scriptsize ${A_{4,4}}$}&
{\scriptsize$[X_1,X_4]=X_1,[X_2,X_4]=X_1+X_2,$}&{\scriptsize$X_1,X_2,X_3$}&
{\scriptsize$dx_1X_1e^{x_4}+dx_2(X_2+x_4X_1)e^{x_4}+dx_3(X_3+X_2x_4$}\\

\vspace{1mm}

{\scriptsize $$}&{\scriptsize$[X_3,X_4]=X_2+X_3$}&{\scriptsize$$}&{\scriptsize$+X_1x_4^2/2)e^{x_4}+dx_4X_4$}\\

\vspace{1mm}

{\scriptsize ${{A^{a,b}}_{4,5}}$}&
{\scriptsize$[X_1,X_4]=X_1,[X_2,X_4]=aX_2,[X_3,X_4]=bX_3$}&{\scriptsize$X_1,X_2,X_3$}&
{\scriptsize $dx_1X_1e^{x_4}+dx_2X_2e^{ax_4}+dx_3X_3e^{bx_4}+dx_4X_4$}\\

\vspace{1mm}

{\scriptsize ${{A^{a,a}}_{4,5}}$}&
{\scriptsize$[X_1,X_4]=X_1,[X_2,X_4]=aX_2,[X_3,X_4]=aX_3$}&{\scriptsize$X_1,X_2,X_3$}&
{\scriptsize $dx_1X_1e^{x_4}+dx_2X_2e^{ax_4}+dx_3X_3e^{ax_4}+dx_4X_4$}\\

\vspace{1mm}

{\scriptsize ${{A^{a,1}}_{4,5}}$}&
{\scriptsize$[X_1,X_4]=X_1,[X_2,X_4]=aX_2,[X_3,X_4]=X_3$}&{\scriptsize$X_1,X_2,X_3$}&
{\scriptsize $dx_1X_1e^{x_4}+dx_2X_2e^{ax_4}+dx_3X_3e^{x_4}+dx_4X_4$}\\

\vspace{1mm}

{\scriptsize ${{A^{1,1}}_{4,5}}$}&
{\scriptsize$[X_1,X_4]=X_1,[X_2,X_4]=X_2,[X_3,X_4]=X_3$}&{\scriptsize$X_1,X_2,X_3$}&
{\scriptsize $dx_1X_1e^{x_4}+dx_2X_2e^{x_4}+dx_3X_3e^{x_4}+dx_4X_4$}\\

\vspace{-1mm}

{\scriptsize ${{A^{a,b}}_{4,6}}$}&
{\scriptsize$[X_1,X_4]=aX_1,[X_2,X_4]=2X_2-X_3,$}&{\scriptsize$X_1,X_2,X_3$}&

{\scriptsize $dx_1X_1e^{ax_4}+e^{bx_4}[dx_2(X_2\cos x_4-X_3\sin x_4)$}\\

\vspace{-1mm}

{\scriptsize $$}&{\scriptsize$[X_3,X_4]=X_2+bX_3$}&{\scriptsize
$$}&{\scriptsize$+dx_3(X_3\cos x_4+X_2\sin
x_4)]+dx_4X_4$}\\\smallskip \\

\hline\hline
    \end{tabular}
\end{center}
\newpage

\vspace{4mm}
\begin{center}
\hspace{10mm}{\small {\bf Table 1}} : Continue \hspace{-1mm}\\
    \begin{tabular}{l l l l  l l p{25mm} }
    \hline\hline
   \vspace{-1mm}
{\scriptsize ${\bf Algebra}$ }& {\scriptsize Non-zero commutation
relations }&{\scriptsize Isometry }&{\scriptsize $g^{-1}
dg$}\\\hline

\vspace{-1mm}

{\scriptsize ${{A}_{4,7}}$}&
{\scriptsize$[X_1,X_4]=2X_1,[X_2,X_4]=X_2,$}&{\scriptsize$X_1,X_2$}&

{\scriptsize $dx_1X_1e^{2x_4}+dx_2(X_2e^{x_4}-x_3X_1e^{2x_4})+dx_3e^{x_4}(x_4X_2$}\\

\vspace{1mm}

{\scriptsize$$}&{\scriptsize$[X_3,X_4]=X_2+X_3,[X_2,X_3]=X_1$}&{\scriptsize$$}&{\scriptsize$+X_3)+dx_4X_4$}\\

\vspace{1mm}

{\scriptsize ${{A}_{4,8}}$}&
{\scriptsize$[X_2,X_4]=X_2,[X_3,X_4]=-X_3,[X_2,X_3]=X_1$}&{\scriptsize$X_1,X_2$}&
{\scriptsize $dx_1X_1+dx_2(X_2e^{x_4}+x_3X_1)+dx_3e^{-x_4}X_3+dx_4X_4$}\\

\vspace{-1mm}

{\scriptsize ${{A}^b_{4,9}}$}&
{\scriptsize$[X_1,X_4]=(1+b)X_1,[X_2,X_4]=X_2,$}&{\scriptsize$X_1,X_2$}&
{\scriptsize $dx_1X_1e^{(1+b)x_4}+dx_2(X_2e^{x_4}+x_3e^{(1+b)x_4}X_1)$}\\

\vspace{1mm}

{\scriptsize
$$}&{\scriptsize$[X_3,X_4]=bX_3,[X_2,X_3]=X_1$}&{\scriptsize $$}&
{\scriptsize$+dx_3e^{x_4}(x_4X_2+X_3)+dx_4X_4$}\\

\vspace{-1mm}

{\scriptsize${{A}^1_{4,9}}$}&{\scriptsize$[X_1,X_4]=2X_1,[X_2,X_4]=X_2,[X_3,X_4]=X_3$},&{\scriptsize$X_1,X_2$}&

{\scriptsize$dx_1X_1e^{2x_4}+dx_2(X_2e^{x_4}+x_3e^{2x_4}X_1)+dx_3e^{x_4}X_3+dx_4X_4$}\\

\vspace{1mm}

& {\scriptsize$[X_2,X_3]=X_1$}\\

\vspace{2mm}

{\scriptsize${{A^0}_{4,9}}$}&{\scriptsize$[X_1,X_4]=X_1,[X_2,X_4]=X_2,[X_2,X_3]=X_1$}&
{\scriptsize$X_1,X_2$}& {\scriptsize $dx_1X_1e^{x_4}+dx_2(X_2e^{x_4}+x_3e^{x_4}X_1)+dx_3X_3+dx_4X_4$}\\

\vspace{-1mm}

{\scriptsize$A_{4,10}$}&{\scriptsize$[X_2,X_4]=-X_3,\;[X_3,X_4]=X_2,;[X_2,X_3]=X_1$}&
{\scriptsize $X_1,X_2$} & {\scriptsize $dx_1X_1+dx_2(x_3X_1+X_2\cos{x_4}-X_3\sin{x_4})+dx_3(X_3\cos{x_4}$}\\

\vspace{1mm}

& &  & {\scriptsize $+X_2\sin{x_4})+dx_4X_4$}\\

\vspace{-1mm}

{\scriptsize${{A^a}_{4,11}}$}&{\scriptsize$[X_1,X_4]=2aX_1,[X_2,X_4]=aX_2-X_3,$}&{\scriptsize$X_1,X_2$}&

{\scriptsize $dx_1X_1e^{2ax_4}+dx_2(x_3e^{2ax_4}X_1+X_2e^{ax_4}\cos{x_4}-$}\\

\vspace{1mm}

{\scriptsize$$}&{\scriptsize$[X_3,X_4]=X_2+aX_3,[X_2,X_3]=X_1$}&{\scriptsize$$}&{\scriptsize$X_3e^{ax_4}\sin{x_4})+dx_3(X_3e^{ax_4}\cos{x_4}+X_2e^{ax_4}\sin{x_4})+dx_4X_4$}\\

\vspace{-1mm}

{\scriptsize ${A_{4,12}}$}&
{\scriptsize$[X_1,X_4]=-X_2,[X_1,X_3]=X_1,[X_2,X_4]=X_1,$}&{\scriptsize$X_1,X_2$}&

{\scriptsize $dx_1e^{x_3}(X_1\cos{x_4}-X_2\sin{x_4})+dx_2e^{x_3}(X_2\cos{x_4}+X_1\sin{x_4})$}\\

\vspace{-1mm}

& {\scriptsize$[X_2,X_3]=X_2$}&{\scriptsize$$}&{\scriptsize$+dx_3X_3+dx_4X_4$}\smallskip \\

\hline\hline
\end{tabular}
\end{center}

\section{\bf Exact solutions}
Here, we construct the $(i,i)$ component of equation (2) for metrics
over all real four dimensional Lie groups of the table (1), then by
comparing these equations with (9) we obtain potentials $V_i$,
furthermore we try to solve equations (18) and obtain initial value
equation (19).
\subsection{\bf Lie algebras $I\oplus R$}
We obtain the potentials $V_i$ as follows:
\begin{eqnarray}
&&\hspace{-110mm}V_1=V_2=V_3=V_4=0,
\end{eqnarray}
such that there are no constraint equations. Equations (18) have the
following form:
\begin{eqnarray}
&&\hspace{-105mm}{\ln(a_i^2 e^{\phi})}''=0, \;\;\; i= 1,2,...,4
\end{eqnarray}
which admit the following general solution:
\begin{eqnarray}
&&\hspace{-135mm}a_i^2e^{\phi}=L_ie^{p_i\tau},
\end{eqnarray}
where $p_i$ and $L_i (i=1,2,...,4)$ are constants; furthermore from
the initial value equation (19) we have the following restriction on
$p_{i} s$:
\begin{eqnarray}
&&\hspace{-125mm}\sum_{i<j}p_ip_j=N\sum_{i}p_i.
\end{eqnarray}
These results are extension of the results of \cite{Ba}.

\subsection{\bf Lie algebras $III\oplus R$}
We obtain the potentials $V_i$ as follows:
\begin{eqnarray}
&&\hspace{-70mm}V_1=-\frac{a_2^2}{2a_1^2a_3^2}-\frac{a_3^2}{2a_1^2a_2^2}-\frac{3}{a_1^2},\:\:
V_2=\frac{a_2^2}{2a_1^2a_3^2}-\frac{a_3^2}{2a_1^2a_2^2}-\frac{2}{a_1^2},\nonumber\\
&&\hspace{-70mm}V_3=-\frac{a_2^2}{2a_1^2a_3^2}+\frac{a_3^2}{2a_1^2a_2^2}-\frac{2}{a_1^2},\:\:V_4=0.
\end{eqnarray}
In this case the constraints imposed by (10) and (11) are
\begin{eqnarray}
&&\hspace{-110mm}a_2a_3=B a_1^2,\nonumber\\
&&\hspace{-110mm}a_2^2+a_3^2=0,
\end{eqnarray}
where $B$ is a constant. The above equations are inconsistent; and
there is no solution for this example.

\subsection{\bf Lie algebras $A_2\bigoplus A_2$}
We obtain the potentials $V_i$ as follows:
\begin{eqnarray}
&&\hspace{-50mm}V_1=V_2=\frac{1}{a_1^2},\:\:V_3=V_4=\frac{1}{a_3^2}.
\end{eqnarray}
The constraint equations (10) impose in this case the restriction
\begin{eqnarray}
&&\hspace{-50mm}\frac{a_1}{a_2}=B_1,\:\frac{a_3}{a_4}=B_2,
\end{eqnarray}
where $B_1$ and $B_2$ are constants. Equations (18) have the
following forms
\begin{eqnarray}
&&\hspace{-70mm}{\ln(a_1^2 e^{\phi})}''-2a_2^2a_3^2a_4^2e^{2\phi}=0,\nonumber\\
&&\hspace{-70mm}{\ln(a_2^2e^{\phi})}''-2a_2^2a_3^2a_4^2e^{2\phi}=0,\nonumber\\
&&\hspace{-70mm}{\ln(a_3^2e^{\phi})}''-2a_1^2a_2^2a_4^2e^{2\phi}=0,\nonumber\\
&&\hspace{-70mm}{\ln(a_4^2 e^{\phi})}''-2a_1^2a_2^2a_4^2e^{2\phi}=0,
\end{eqnarray}
which admit the following general solution:
\begin{eqnarray}
&&\hspace{-70mm}a_1^2e^{\phi}=a_3^2e^{\phi}=p_4^2\left(\frac{(B_1B_2)^2e^{N\tau}}{3\cosh^2(p_4\tau)}\right)^{\frac{1}{3}},\nonumber\\
&&\hspace{-70mm}a_2^2e^{\phi}=p_4^2\left(\frac{(\frac{B_2}{B_1^2})^2e^{N\tau}}{3\cosh^2(p_4\tau)}\right)^{\frac{1}{3}},\nonumber\\
&&\hspace{-70mm}a_4^2e^{\phi}=p_4^2\left(\frac{(\frac{B_1}{B_2^2})^2e^{N\tau}}{3\cosh^2(p_4\tau)}\right)^{\frac{1}{3}},
\end{eqnarray}
where $p_4$ is constant. From the initial value equation (19) we
have the following restriction on $p_4$:
\begin{eqnarray}
&&\hspace{-70mm}N^2-4B_1^2B_2^2p_4^6=0.
\end{eqnarray}

\subsection{\bf Lie algebras $II\bigoplus R$}
We obtain the potentials $V_i$ as follows:
\begin{eqnarray}
&&\hspace{-50mm}V_1=-V_2=-V_3=\frac{a_1^2}{2a_2^2a_3^2},\;\: V_4=0,
\end{eqnarray}
such that there are no constraint equations. Equations (18) have the
following form:
\begin{eqnarray}
&&\hspace{-70mm}{\ln(a_1^2 e^{\phi})}''+a_1^4a_4^2e^{2\phi}=0,\nonumber\\
&&\hspace{-70mm}{\ln(a_2^2e^{\phi})}''-a_1^4a_4^2e^{2\phi}=0,\nonumber\\
&&\hspace{-70mm}{\ln(a_3^2e^{\phi})}''-a_1^4a_4^2e^{2\phi}=0,\nonumber\\
&&\hspace{-70mm}{\ln(a_4^2 e^{\phi})}''=0,
\end{eqnarray}
which admit the following general solution:
\begin{eqnarray}
&&\hspace{-100mm}a_1^2e^{\phi}=\frac{p_1^2\:e^{(\frac{N-p_4}{2})\tau}}{\sqrt{L_1}\cosh(p_1\tau)},\nonumber\\
&&\hspace{-100mm}a_2^2e^{\phi}=L_2^2\cosh(p_1\tau)e^{p_2\tau},\nonumber\\
&&\hspace{-100mm}a_3^2e^{\phi}=L_3^2\cosh(p_1\tau)e^{p_3\tau},\nonumber\\
&&\hspace{-100mm}a_4^2e^{\phi}=L_1e^{p_4\tau},
\end{eqnarray}
where $p_1,..., p_4, L_1, L_2$ and $L_3$ are constants. Note that
these solutions are extension of solutions \cite{Ba} for Bianchi
algebras $II$. From the initial value equation (19) we have the
following restriction on $p_{i}s$:
\begin{eqnarray}
&&\hspace{-70mm}N(p_2+p_3+N)+p_4(p_2+p_3+p_4)-2p_2p_3+2p_1^2=0.
\end{eqnarray}

\subsection{\bf Lie algebras ($IV\oplus R$)}
We obtain the potentials $V_i$ as follows:
\begin{eqnarray}
&&\hspace{-70mm}V_1=V_2=-\frac{a_3^2}{2a_1^2a_2^2}-\frac{2}{a_1^2},\:\:V_3=\frac{a_3^2}{2a_1^2a_2^2}-\frac{2}{a_1^2},\:\:V_4=0.
\end{eqnarray}
In this case the constraints imposed by (10) and (11) are as
follows:
\begin{eqnarray}
&&\hspace{-110mm}a_2a_3=B a_1^2,\nonumber\\
&&\hspace{-110mm}a_3=0,
\end{eqnarray}
where $B$ is constant. There is no solution for these inconsistent
constraints.

\subsection{Lie algebras $A_{3,3}\oplus A_{1}\approx V \oplus R$}
In this case we obtain the potentials $V_i$ as follows:
\begin{eqnarray}
&&\hspace{-50mm}V_1=V_2=V_3=\frac{2}{a_1^2},\:\; V_4=0,
\end{eqnarray}
and the constraint imposed by (10) and (11) is
\begin{eqnarray}
&&\hspace{-110mm}a_2a_3=B a_1^2.
\end{eqnarray}
Equations (18) have the following forms:
\begin{eqnarray}
&&\hspace{-70mm}{\ln(a_1^2 e^{\phi})}''-4a_2^2a_3^2a_4^2e^{2\phi}=0,\nonumber\\
&&\hspace{-70mm}{\ln(a_2^2e^{\phi})}''-4a_2^2a_3^2a_4^2e^{2\phi}=0,\nonumber\\
&&\hspace{-70mm}{\ln(a_3^2e^{\phi})}''-4a_2^2a_3^2a_4^2e^{2\phi}=0,\nonumber\\
&&\hspace{-70mm}{\ln(a_4^2 e^{\phi})}''=0,
\end{eqnarray}
which admit the following general solution:
\begin{eqnarray}
&&\hspace{-100mm}a_1^2e^{\phi}=\frac{p_1\:e^{(\frac{N-p_4}{2})\tau}}{2B\sqrt{L_1}\sinh(p_1\tau)},\nonumber\\
&&\hspace{-100mm}a_2^2e^{\phi}=\frac{L_2\:e^{p_2\tau}}{2\sinh(p_1\tau)},\nonumber\\
&&\hspace{-100mm}a_3^2e^{\phi}=\frac{p_1^2\:e^{(N-p_2-p_4)\tau}}{2L_1L_2\sinh(p_1\tau)},\nonumber\\
&&\hspace{-100mm}a_4^2e^{\phi}=L_1e^{p_4\tau},
\end{eqnarray}
where $p_1, p_2, p_4, L_1$ and $L_2$ are constants.  From the
initial value equation (19) we have the following restriction on
$p_{i}s$:
\begin{eqnarray}
&&\hspace{-70mm}N(p_2+p_4-N)-p_2(p_2+p_4)-\left(p_4^2-3p_1^2\right)=0.
\end{eqnarray}

\subsection{\bf Lie Algebras $A_{3,4}\bigoplus A_{1}\approx VI_{0} \bigoplus R$}
We obtain the potentials $V_i$ as follows:
\begin{eqnarray}
&&\hspace{-50mm}V_1=-V_2=\frac{a_1^2}{2a_2^2a_3^2}-\frac{a_2^2}{2a_1^2a_3^2},\:\;
V_3=-\frac{a_1^2}{2a_2^2a_3^2}-\frac{a_2^2}{2a_1^2a_3^2}-\frac{1}{a_3^2},\:\;
V_4=0,
\end{eqnarray}
such that there are no constraint equations. Equations (18) have
the following forms:
\begin{eqnarray}
&&\hspace{-70mm}{\ln(a_1^2 e^{\phi})}''+(a_1^4-a_2^4)a_4^2e^{2\phi}=0,\nonumber\\
&&\hspace{-70mm}{\ln(a_2^2e^{\phi})}''-(a_1^4-a_2^4)a_4^2e^{2\phi}=0,\nonumber\\
&&\hspace{-70mm}{\ln(a_3^2e^{\phi})}''-(a_1^4+a_2^4)a_4^2e^{2\phi}-2a_1^2a_2^2a_4^2e^{2\phi}=0,\nonumber\\
&&\hspace{-70mm}{\ln(a_4^2 e^{\phi})}''=0,
\end{eqnarray}
which admit the following general solution:
\begin{eqnarray}
&&\hspace{-80mm}a_1^2 e^{\phi}=a_2^2e^{\phi}=\sqrt{L_1}\:e^{\frac{p_1}{2}\tau},\nonumber\\
&&\hspace{-80mm}a_3^2e^{\phi}=exp\left(\frac{4L_1L_2}{(p_1+p_4-N)^2}\,e^{(p_1+p_4-N)\tau}\right),\nonumber\\
&&\hspace{-80mm}a_4^2e^{\phi}=L_2e^{p_4\tau},
\end{eqnarray}
where $L_1, L_2, p_1$ and $p_4$ are constants. From the initial
value equation (19) we have the following restriction on $p_1$ and
$p_4$:
\begin{eqnarray}
&&\hspace{-100mm}p_1(p_1+4p_4-4N)-4p_4N=0.
\end{eqnarray}
Not that it has not presented solution for Bianchi algebra $VI_0$ in
\cite{Ba}
\subsection{\bf Lie Algebras $VI_{b}\oplus R$}
We obtain the potentials $V_i$ as follows:
\begin{eqnarray}
&&\hspace{-50mm}V_1=-\frac{a_2^2}{2a_1^2a_3^2}-\frac{a_3^2}{2a_1^2a_2^2}-\frac{1+2b^2}{a_1^2},\:\:
V_2=\frac{a_2^2}{2a_1^2a_3^2}-\frac{a_3^2}{2a_1^2a_2^2}-\frac{2b^2}{a_1^2},\nonumber\\
&&\hspace{-50mm}V_3=-\frac{a_2^2}{2a_1^2a_3^2}+\frac{a_3^2}{2a_1^2a_2^2}-\frac{2b^2}{a_1^2},\:\:
V_4=0.
\end{eqnarray}
In this case, the constraints imposed by (10) and (11) are
\begin{eqnarray}
&&\hspace{-110mm}a_2a_3=B a_1^2,\nonumber\\
&&\hspace{-110mm}a_2^2+a_3^2=0,
\end{eqnarray}
where $B$ is constant; so that, within the present context, all
solutions are singular.
\subsection{Lie algebras $VII_0\bigoplus R$}
We obtain the potentials $V_i$ as follows:
\begin{eqnarray}
&&\hspace{-50mm}V_1=-V_2=\frac{a_1^2}{2a_2^2a_3^2}-\frac{a_2^2}{2a_1^2a_3^2},\:\;
V_3=-\frac{a_1^2}{2a_2^2a_3^2}-\frac{a_2^2}{2a_1^2a_3^2}+\frac{1}{a_3^2},\:\;
V_4=0,
\end{eqnarray}
such that there are no constraint equations. Equations (18) have
the following forms:
\begin{eqnarray}
&&\hspace{-70mm}{\ln(a_1^2 e^{\phi})}''+(a_1^4-a_2^4)a_4^2e^{2\phi}=0,\nonumber\\
&&\hspace{-70mm}{\ln(a_2^2e^{\phi})}''-(a_1^4-a_2^4)a_4^2e^{2\phi}=0,\nonumber\\
&&\hspace{-70mm}{\ln(a_3^2e^{\phi})}''-(a_1^4+a_2^4)a_4^2e^{2\phi}+2a_1^2a_2^2a_4^2e^{2\phi}=0,\nonumber\\
&&\hspace{-70mm}{\ln(a_4^2 e^{\phi})}''=0,
\end{eqnarray}
which admit the following general solution
\begin{eqnarray}
&&\hspace{-110mm}a_1^2 e^{\phi}=a_2^2e^{\phi}=\sqrt{L_1}\:e^{\frac{p_1}{2}\tau},\nonumber\\
&&\hspace{-110mm}a_3^2e^{\phi}=L_2e^{p_3\tau},\nonumber\\
&&\hspace{-110mm}a_4^2e^{\phi}=L_3e^{p_4\tau},
\end{eqnarray}
where $L_1, L_2, L_3, p_1, p_3$ and $p_4$ are constants. Furthermore
from the initial value equation (19) we have the following
restriction on $p_{i}s$:
\begin{eqnarray}
&&\hspace{-70mm}\frac{p_1^2}{4}+p_1p_3+p_1p_4+p_3p_4-\left(p_1+p_3+p_4\right)N=0.
\end{eqnarray}
Note that this solution and initial value equation are the same as
for the Lie algebra $I\oplus R$.

\subsection{\bf Lie algebras $VII_b\oplus R$}
We obtain the potentials $V_i$ as follows:
\begin{eqnarray}
&&\hspace{-50mm}V_1=-\frac{a_3^2}{2a_1^2a_2^2}-\frac{a_2^2}{2a_1^2a_3^2}-\frac{2b^2-1}{a_1^2},\:\;
V_2=\frac{a_2^2}{2a_1^2a_3^2}-\frac{a_3^2}{2a_1^2a_2^2}-\frac{2b^2}{a_1^2},\nonumber\\
&&\hspace{-50mm}V_3=-\frac{a_2^2}{2a_1^2a_3^2}+\frac{a_3^2}{2a_1^2a_2^2}-\frac{2b^2}{2a_1^2},\:\;
V_4=0.
\end{eqnarray}
In this case the constraints imposed by (10), (11) are
\begin{eqnarray}
&&\hspace{-110mm}a_2a_3=B a_1^2,\nonumber\\
&&\hspace{-110mm}a_2^2-a_3^2=0,
\end{eqnarray}
where $B$ is constant. Equations (18) have the following forms:
\begin{eqnarray}
&&\hspace{-70mm}{\ln(a_1^2 e^{\phi})}''-\left((a_2^2-a_3^2)^2+4b^2a_2^2a_3^2\right)a_4^2e^{2\phi}=0,\nonumber\\
&&\hspace{-70mm}{\ln(a_2^2e^{\phi})}''-4b^2a_2^2a_3^2a_4^2e^{2\phi}=0,\nonumber\\
&&\hspace{-70mm}{\ln(a_3^2e^{\phi})}''-4b^2a_2^2a_3^2a_4^2e^{2\phi}=0,\nonumber\\
&&\hspace{-70mm}{\ln(a_4^2 e^{\phi})}''=0,
\end{eqnarray}
which admit the following general solution:
\begin{eqnarray}
&&\hspace{-100mm}a_1^2e^{\phi}=\frac{p_1\:e^{(\frac{N-p_4}{2})\tau}}{2bB\sqrt{L_1}\sinh(p_1\tau)},\nonumber\\
&&\hspace{-100mm}a_2^2e^{\phi}=a_3^2e^{\phi}=\frac{p_1\:e^{(\frac{N-p_4}{2})\tau}}{2b\sqrt{L_1}\sinh(p_1\tau)},\nonumber\\
&&\hspace{-100mm}a_4^2e^{\phi}=L_1e^{p_4\tau},
\end{eqnarray}
where $L_1, p_1$ and $p_4$ are constants. From the initial value
equation (19) we have the following restriction on $p_1$ and $p_4$:
\begin{eqnarray}
&&\hspace{-100mm}4p_1^2-p_4^2-N^2+\frac{2}{3}Np_4=0.
\end{eqnarray}

\subsection{\bf Lie algebras $VIII\oplus R$}
We obtain the potentials $V_i$ as follows:
\begin{eqnarray}
&&\hspace{-50mm}V_1=\frac{a_1^2}{2a_2^2a_3^2}-\frac{a_2^2}{2a_1^2a_3^2}-\frac{2a_2^2+a_3^2}{2a_1^2a_2^2},\:\;
V_2=-\frac{a_1^2}{2a_2^2a_3^2}+\frac{a_2^2}{2a_1^2a_3^2}-\frac{2a_1^2+a_3^2}{2a_1^2a_2^2},\nonumber\\
&&\hspace{-50mm}V_3=-\frac{a_2^2}{2a_1^2a_3^2}+\frac{a_3^2}{2a_1^2a_2^2}-\frac{a_1^2-2a_2^2}{2a_2^2a_3^2},\:\;
V_4=0,
\end{eqnarray}
such that there are no constraint equations. Equations (18) have
the following forms:
\begin{eqnarray}
&&\hspace{-70mm}{\ln(a_1^2 e^{\phi})}''+\left(a_1^4-(a_2^2+a_3^2)^2\right) a_4^2e^{2\phi}=0,\nonumber\\
&&\hspace{-70mm}{\ln(a_2^2e^{\phi})}''+\left(a_2^4-(a_1^2+a_3^2)^2\right) a_4^2e^{2\phi}=0,\nonumber\\
&&\hspace{-70mm}{\ln(a_3^2e^{\phi})}''+\left(a_3^4-(a_1^2-a_2^2)^2\right) a_4^2e^{2\phi}=0,\nonumber\\
&&\hspace{-70mm}{\ln(a_4^2 e^{\phi})}''=0,
\end{eqnarray}
which admit the following general solution:
\begin{eqnarray}
&&\hspace{-100mm}a_1^2e^{\phi}=a_2^2e^{\phi}=\frac{p_1^2\:e^{(\frac{N-p_4}{2})\tau}\cosh(p_3\tau)}{\sqrt{L_1}p_3\sinh^2(p_1\tau)},\nonumber\\
&&\hspace{-100mm}a_3^2e^{\phi}=\frac{p_3\:e^{(\frac{N-p_4}{2})\tau}}{\sqrt{L_1}\cosh(p_3\tau)},\nonumber\\
&&\hspace{-100mm}a_4^2e^{\phi}=L_1e^{p_4\tau},
\end{eqnarray}
where $L_1, p_1, p_3$ and $p_4$ are constants. From the initial
value equation (19) we have the following restriction on $p_{i}s$:
\begin{eqnarray}
&&\hspace{-100mm}4p_1^2-p_3^2-\frac{3}{4}p_4^2+\frac{5}{4}Np_4-\frac{3}{4}N^2=0.
\end{eqnarray}

\subsection{\bf Lie algebras $IX\oplus R$}
We obtain the potentials $V_i$ as follows:
\begin{eqnarray}
&&\hspace{-50mm}V_1=\frac{a_1^2}{2a_2^2a_3^2}-\frac{a_2^2}{2a_1^2a_3^2}+\frac{2a_2^2-a_3^2}{2a_1^2a_2^2},\:\;
V_2=-\frac{a_1^2}{2a_2^2a_3^2}+\frac{a_2^2}{2a_1^2a_3^2}+\frac{2a_1^2-a_3^2}{2a_1^2a_2^2},\nonumber\\
&&\hspace{-50mm}V_3=-\frac{a_2^2}{2a_1^2a_3^2}+\frac{a_3^2}{2a_1^2a_2^2}-\frac{a_1^2-2a_2^2}{2a_2^2a_3^2},\:\;
V_4=0,
\end{eqnarray}
such that there are no constraint equations. Equations (18) have
the following forms:
\begin{eqnarray}
&&\hspace{-70mm}{\ln(a_1^2 e^{\phi})}''+\left(a_1^4-(a_2^2-a_3^2)^2\right) a_4^2e^{2\phi}=0,\nonumber\\
&&\hspace{-70mm}{\ln(a_2^2e^{\phi})}''+\left(a_2^4-(a_1^2-a_3^2)^2\right) a_4^2e^{2\phi}=0,\nonumber\\
&&\hspace{-70mm}{\ln(a_3^2e^{\phi})}''+\left(a_3^4-(a_1^2-a_2^2)^2\right) a_4^2e^{2\phi}=0,\nonumber\\
&&\hspace{-70mm}{\ln(a_4^2 e^{\phi})}''=0,
\end{eqnarray}
which admit the following general solution:
\begin{eqnarray}
&&\hspace{-100mm}a_1^2e^{\phi}=a_2^2e^{\phi}=\frac{p_1^2\:e^{(\frac{N-p_4}{2})\tau}\cosh(p_3\tau)}{\sqrt{L_1}p_3\cosh^2(p_1\tau)},\nonumber\\
&&\hspace{-100mm}a_3^2e^{\phi}=\frac{p_3\:e^{(\frac{N-p_4}{2})\tau}}{\sqrt{L_1}\cosh(p_3\tau)},\nonumber\\
&&\hspace{-100mm}a_4^2e^{\phi}=L_1e^{p_4\tau},
\end{eqnarray}
where $L_1, p_1, p_3$ and $p_4$ are constants. From the initial
value equation (19) we have the following restriction on $p_{i}s$:
\begin{eqnarray}
&&\hspace{-100mm}4p_1^2-p_3^2-\frac{3}{4}p_4^2+\frac{5}{4}Np_4-\frac{3}{4}N^2=0.
\end{eqnarray}
Note that equations (63) and (67) for $p_4=0$ are the same initial
values equations of the Ref \cite{Ba} for Lie algebras $VIII$ and
$IX$ respectively such that the scalar field play the role of
antisymmetric field i.e. $\frac{3}{4}N^2=A^2$; where $A$ is a
parameter of antisymmetric field \cite{Ba}.

\subsection{\bf Lie algebras $A_{4,1}$}
We obtain the potentials $V_i$ as follows:
\begin{eqnarray}
&&\hspace{-30mm}V_1=\frac{a_1^2}{2a_2^2a_4^2},\:\:V_2=-\frac{a_1^2}{2a_2^2a_4^2}+\frac{a_2^2}{2a_3^2a_4^2},\:\:
V_3=-\frac{a_2^2}{2a_3^2a_4^2},\:\:V_4=-\frac{a_1^2}{2a_2^2a_4^2}-\frac{a_2^2}{2a_3^2a_4^2},
\end{eqnarray}
such that there are no constraint equations. Equations (18) have
the following forms:
\begin{eqnarray}
&&\hspace{-70mm}{\ln(a_1^2 e^{\phi})}''+a_1^4a_3^2e^{2\phi}=0,\nonumber\\
&&\hspace{-70mm}{\ln(a_2^2e^{\phi})}''+\left(a_1^2a_2^4-a_3^2a_1^4\right)e^{2\phi}=0,\nonumber\\
&&\hspace{-70mm}{\ln(a_3^2e^{\phi})}''-a_1^2a_2^4e^{2\phi}=0,\nonumber\\
&&\hspace{-70mm}{\ln(a_4^2e^{\phi})}''-\left(a_1^2a_2^4+a_3^2a_1^4\right)e^{2\phi}=0,
\end{eqnarray}
which admit the following general solution:
\begin{eqnarray}
&&\hspace{-70mm}a_1^2e^{\phi}=\frac{2p_2^{2}}{L_2^2\cosh^2(p_2\tau)}\,e^{(N-2p_2)\tau},\nonumber\\
&&\hspace{-70mm}a_2^2e^{\phi}=L_2\,e^{p_2\tau},\nonumber\\
&&\hspace{-70mm}a_3^2e^{\phi}=\frac{L_2^4\cosh^2(p_2\tau)}{2p_2^2}\,e^{(4p_2-N)\tau},\nonumber\\
&&\hspace{-70mm}a_4^2e^{\phi}=L_4\cosh^4(p_2\tau)e^{p_4\tau},
\end{eqnarray}
where $L_2, L_4, p_2$ and $p_4$ are constants. From the initial
value equation (19) we have the following restriction on $p_2$ and
$p_4$:
\begin{eqnarray}
&&\hspace{-70mm}N\left(N+p_4-3p_2\right)+7p_2p_4=0.
\end{eqnarray}

\subsection{\bf Lie algebras ($A_{4,2}^{b}$)}
We obtain the potentials $V_i$ as follows:
\begin{eqnarray}
&&\hspace{-70mm}V_1=-\frac{2+b^2}{a_4^2},\:\:V_2=\frac{a_2^2}{2a_3^2a_4^2}-\frac{2+b}{a_4^2},\:\:\nonumber\\
&&\hspace{-70mm}V_3=-\frac{a_2^2}{2a_3^2a_4^2}-\frac{2+b}{a_4^2},\:\:V_4=-\frac{a_2^2}{2a_3^2a_4^2}-\frac{2+b^2}{a_4^2}.
\end{eqnarray}
In this case the constraints imposed by (10) and (11) are
\begin{eqnarray}
&&\hspace{-110mm}a_1^ba_2a_3=B a_4^{b+2},\nonumber\\
&&\hspace{-110mm}a_2=0,
\end{eqnarray}
where $B$ is constant; so that, within the present context, all
solutions are singular.
\subsection{\bf Lie algebras ($A_{4,2}^{1}$)}
We obtain the potentials $V_i$ as follows:
\begin{eqnarray}
&&\hspace{-90mm}V_1=-\frac{3}{a_4^2},\:\:V_2=\frac{a_2^2}{2a_3^2a_4^2}-\frac{3}{a_4^2},\:\nonumber\\
&&\hspace{-90mm}V_3=-\frac{a_2^2}{2a_3^2a_4^2}-\frac{3}{a_4^2},\:\:V_4=-\frac{a_2^2}{2a_3^2a_4^2}-\frac{3}{a_4^2}.
\end{eqnarray}
In this case the constraints imposed by (10) and (11) are
\begin{eqnarray}
&&\hspace{-110mm}a_1a_2a_3=B a_4^{3},\nonumber\\
&&\hspace{-110mm}a_2=0,
\end{eqnarray}
where $B$ is constant; so that, within the present context, all
solutions are singular.
\subsection{\bf Lie algebras ($A_{4,3}$)}
We obtain the potentials $V_i$ as follows:
\begin{eqnarray}
&&\hspace{-70mm}V_1=-\frac{1}{a_4^2},\:\:V_2=\frac{a_2^2}{2a_3^2a_4^2},\:\:V_3=-\frac{a_2^2}{2a_3^2a_4^2},\:\:
V_4=-\frac{a_2^2}{2a_3^2a_4^2}-\frac{1}{a_4^2}.
\end{eqnarray}
In this case the constraints imposed by (10) and (11) are
\begin{eqnarray}
&&\hspace{-110mm}a_1=B a_4,\nonumber\\
&&\hspace{-110mm}a_2=0,
\end{eqnarray}
where $B$ is constant; so that, within the present context, all
solutions are singular.

\subsection{\bf Lie algebras $A_{4,4}$}
We obtain the potentials $V_i$ as follows:
\begin{eqnarray}
&&\hspace{-70mm}V_1=\frac{a_1^2}{2a_2^2a_4^2}-\frac{3}{a_4^2},\:\:V_2=-\frac{a_2^2}{2a_3^2a_4^2}-\frac{3}{a_4^2},\nonumber\\
&&\hspace{-70mm}V_3=-\frac{a_1^2}{2a_2^2a_4^2}+\frac{a_2^2}{2a_3^2a_4^2}-\frac{3}{a_4^2},\:\:
V_4=-\frac{a_1^2}{2a_2^2a_4^2}-\frac{a_2^2}{2a_3^2a_4^2}-\frac{3}{a_4^2}.
\end{eqnarray}
In this case the constraints imposed by (10) and (11) are
\begin{eqnarray}
&&\hspace{-110mm}a_1a_2a_3=B a_4^3,\nonumber\\
&&\hspace{-110mm}a_1=0,
\end{eqnarray}
where $B$ is constant; so that, within the present context, all
solutions are singular.

\subsection{\bf Lie algebras $A_{4,5}^{(a,b)}$}
We obtain the potentials $V_i$ as follows:
\begin{eqnarray}
&&\hspace{-90mm}V_1=-\frac{a+b+1}{a_4^2},\:\:V_2=-\frac{a\left(a+b+1\right)}{a_4^2},\:\nonumber\\
&&\hspace{-90mm}V_3=-\frac{b\left(a+b+1\right)}{a_4^2},\:\:V_4=-\frac{a^2+b^2+1}{a_4^2}.
\end{eqnarray}
In this case the constraint equations (10) impose the following
restriction:
\begin{eqnarray}
&&\hspace{-120mm}a_1a_2^{a}a_3^{b}=a_4^{a+b+1}.
\end{eqnarray}
Equations (18) have the following forms:
\begin{eqnarray}
&&\hspace{-70mm}{\ln(a_1^2 e^{\phi})}''-2\left(a+b+1\right)a_1^2a_2^2a_3^2\:e^{2\phi}=0,\nonumber\\
&&\hspace{-70mm}{\ln(a_2^2e^{\phi})}''-2a\left(a+b+1\right)a_1^2a_2^2a_3^2\:e^{2\phi}=0,\nonumber\\
&&\hspace{-70mm}{\ln(a_3^2e^{\phi})}''-2b\left(a+b+1\right)a_1^2a_2^2a_3^2\:e^{2\phi}=0,\nonumber\\
&&\hspace{-70mm}{\ln(a_4^2e^{\phi})}''-2\left(a^2+b^2+1\right)a_1^2a_2^2a_3^2\:e^{2\phi}=0,
\end{eqnarray}
which for $a+b\neq -1$ admit the following general solution:
\begin{eqnarray}
&&\hspace{-70mm}a_1^2e^{\phi}=\frac{L_1^{b}(\frac{p_4}{a+b+1})^{\frac{2}{a+b+1}}\,e^{p_1\tau}}{\sinh(p_4\tau)^{\frac{2}{a+b+1}}},\nonumber\\
&&\hspace{-70mm}a_2^2e^{\phi}=\frac{L_2^{\frac{b}{a}}(\frac{p_4}{a+b+1})^{\frac{2a}{a+b+1}}\,e^{p_2\tau}}{\sinh(p_4\tau)^{\frac{2a}{a+b+1}}},\nonumber\\
&&\hspace{-70mm}a_3^2e^{\phi}=\frac{(\frac{p_4}{a+b+1})^{\frac{2b}{a+b+1}}\,e^{\left(N-p_1-p_2\right)\tau}}{L_1L_2\sinh(p_4\tau)^{\frac{2b}{a+b+1}}},\nonumber\\
&&\hspace{-70mm}a_4^2e^{\phi}=\frac{(\frac{p_4}{a+b+1})^{\frac{2(a^2+b^2+1)}{\left(a+b+1\right)^2}}\,
e^{\left(\frac{p_2(a-b)+p_1(1-b)+bN}{a+b+1}\right)\tau}}{\sinh(p_4\tau)^{\frac{2\left(a^2+b^2+1\right)}{\left(a+b+1\right)^2}}},
\end{eqnarray}
where $L_1$, $L_2$, $p_1$, $p_2$ and $p_4$ are constants. From the
initial value equation (19) we have the following restriction on
$p_{i}s$:
\begin{eqnarray}
&&\hspace{-50mm}p_1^2+p_2^2+N^2+p_1p_2-N(p_1+p_2)-2\left(1+\frac{a^2+b^2+1}{(a+b+1)^2}\right)p_4^2=0.
\end{eqnarray}
For $a+b=-1$ solution is similar to solution of $VI_0$ model.

\subsection{\bf Lie algebras $A_{4,5}^{(a,a)}$}
We obtain the potentials $V_i$ as follows:
\begin{eqnarray}
&&\hspace{-50mm}V_1=-\frac{2a+1}{a_4^2},\:\:V_2=V_3=aV_1=-\frac{a(2a+1)}{a_4^2},\:\:V_4=-\frac{2a^2+1}{a_4^2}.
\end{eqnarray}
The constraint equations (10) impose the following restriction:
\begin{eqnarray}
&&\hspace{-120mm}a_1a_2^{a}a_3^{a}=a_4^{2a+1}.
\end{eqnarray}
Equations (18) have the following forms:
\begin{eqnarray}
&&\hspace{-70mm}{\ln(a_1^2 e^{\phi})}''-2\left(2a+1\right)a_1^2a_2^2a_3^2\:e^{2\phi}=0,\nonumber\\
&&\hspace{-70mm}{\ln(a_2^2e^{\phi})}''-2a\left(2a+1\right)a_1^2a_2^2a_3^2\:e^{2\phi}=0,\nonumber\\
&&\hspace{-70mm}{\ln(a_3^2e^{\phi})}''-2a\left(2a+1\right)a_1^2a_2^2a_3^2\:e^{2\phi}=0,\nonumber\\
&&\hspace{-70mm}{\ln(a_4^2e^{\phi})}''-2\left(2a^2+1\right)a_1^2a_2^2a_3^2\:e^{2\phi}=0,
\end{eqnarray}
which for $a\neq \frac{-1}{2}$ admit the following general
solution:
\begin{eqnarray}
&&\hspace{-70mm}a_1^2e^{\phi}=\frac{L_1^{a}(\frac{p_4}{2a+1})^{\frac{2}{2a+1}}\,e^{p_1\tau}}{\sinh(p_4\tau)^{\frac{2}{2a+1}}},\nonumber\\
&&\hspace{-70mm}a_2^2e^{\phi}=\frac{L_2(\frac{p_4}{2a+1})^{\frac{2a}{2a+1}}\,e^{p_2\tau}}{\sinh(p_4\tau)^{\frac{2a}{2a+1}}},\nonumber\\
&&\hspace{-70mm}a_3^2e^{\phi}=\frac{(\frac{p_4}{2a+1})^{\frac{2a}{2a+1}}\,e^{\left(N-p_1-p_2\right)\tau}}{L_1L_2\sinh(p_4\tau)^{\frac{2a}{2a+1}}},\nonumber\\
&&\hspace{-70mm}a_4^2e^{\phi}=\frac{(\frac{p_4}{2a+1})^{\frac{2(2a^2+1)}{\left(2a+1\right)^2}}\,
e^{\left(\frac{p_1(1-a)+aN}{2a+1}\right)\tau}}{\sinh(p_4\tau)^{\frac{2\left(2a^2+1\right)}{\left(2a+1\right)^2}}},
\end{eqnarray}
where $L_1$, $L_2$, $p_1$, $p_2$ and $p_4$ are constants. From the
initial value equation (19) we have the following restriction on
$p_{i}s$:
\begin{eqnarray}
&&\hspace{-50mm}p_1^2+p_2^2+N^2+p_1p_2-N(p_1+p_2)-2\left(1+\frac{2a^2+1}{(2a+1)^2}\right)p_4^2=0.
\end{eqnarray}
For $a=\frac{-1}{2}$ solution is similar to solution of $VI_0$
model. Note that for this Lie algebra all equations, constraints
solutions and initial value conditions are the same as for Lie
algebra $A_{4,5}^{(a,b)}$ for $b=a$.
\subsection{\bf Lie algebras $A_{4,5}^{(a,1)}$}
We obtain the potentials $V_i$ as follows:
\begin{eqnarray}
&&\hspace{-50mm}V_1=V_3=-\frac{a+2}{a_4^2},\:\:V_2=aV_1=-\frac{a\left(a+2\right)}{a_4^2},\:\:V_4=-\frac{a^2+2}{a_4^2}.
\end{eqnarray}
The constraint equations (10) impose in this case the restriction
\begin{eqnarray}
&&\hspace{-120mm}a_1a_2^{a}a_3=a_4^{a+2}.
\end{eqnarray}
Equations (18) have the following forms:
\begin{eqnarray}
&&\hspace{-70mm}{\ln(a_1^2 e^{\phi})}''-2\left(a+2\right)a_1^2a_2^2a_3^2\:e^{2\phi}=0,\nonumber\\
&&\hspace{-70mm}{\ln(a_2^2e^{\phi})}''-2a\left(a+2\right)a_1^2a_2^2a_3^2\:e^{2\phi}=0,\nonumber\\
&&\hspace{-70mm}{\ln(a_3^2e^{\phi})}''-2\left(a+2\right)a_1^2a_2^2a_3^2\:e^{2\phi}=0,\nonumber\\
&&\hspace{-70mm}{\ln(a_4^2e^{\phi})}''-2\left(a^2+2\right)a_1^2a_2^2a_3^2\:e^{2\phi}=0,
\end{eqnarray}
which admit the following general solution:
\begin{eqnarray}
&&\hspace{-70mm}a_1^2e^{\phi}=\frac{L_1^{b}(\frac{p_4}{a+2})^{\frac{2}{a+2}}\,e^{p_1\tau}}{\sinh(p_4\tau)^{\frac{2}{a+2}}},\nonumber\\
&&\hspace{-70mm}a_2^2e^{\phi}=\frac{L_2(\frac{p_4}{a+2})^{\frac{2a}{a+2}}\,e^{p_2\tau}}{\sinh(p_4\tau)^{\frac{2a}{a+2}}},\nonumber\\
&&\hspace{-70mm}a_3^2e^{\phi}=\frac{(\frac{p_4}{a+2})^{\frac{2}{a+2}}\,e^{\left(N-p_1-p_2\right)\tau}}{(L_1L_2)^a\sinh(p_4\tau)^{\frac{2}{a+2}}},\nonumber\\
&&\hspace{-70mm}a_4^2e^{\phi}=\frac{(\frac{p_4}{a+2})^{\frac{2(a^2+2)}{\left(a+2\right)^2}}\,
e^{\left(\frac{p_2(a-1)+N}{a+2}\right)\tau}}{\sinh(p_4\tau)^{\frac{2\left(a^2+2\right)}{\left(a+2\right)^2}}},
\end{eqnarray}
where $L_1, L_2, p_1, p_2$ and $p_4$ are constants. From the initial
value equation (19) we have the following restriction on $p_i$:
\begin{eqnarray}
&&\hspace{-50mm}p_1^2+p_2^2+N^2+p_1p_2-N(p_1+p_2)-2\left(1+\frac{a^2+2}{(a+2)^2}\right)p_4^2=0.
\end{eqnarray}
Note that the results for this algebra are same as for Lie algebra
$A_{4,5}^{(a,b)}$ for $b=1$.

\subsection{\bf Lie algebras $A_{4,5}^{(1,1)}$}
We obtain the potentials $V_i$ as follows:
\begin{eqnarray}
&&\hspace{-50mm}V_1=V_2=V_3=V_4=-\frac{3}{a_4^2}.
\end{eqnarray}
The constraint equations (10) impose the following restriction:
\begin{eqnarray}
&&\hspace{-120mm}a_1a_2a_3=a_4^{3}.
\end{eqnarray}
Equations (18) have the following forms:
\begin{eqnarray}
&&\hspace{-90mm}{\ln(a_1^2e^{\phi})}''-6a_1^2a_2^2a_3^2\:e^{2\phi}=0,\nonumber\\
&&\hspace{-90mm}{\ln(a_2^2e^{\phi})}''-6a_1^2a_2^2a_3^2\:e^{2\phi}=0,\nonumber\\
&&\hspace{-90mm}{\ln(a_3^2e^{\phi})}''-6a_1^2a_2^2a_3^2\:e^{2\phi}=0,\nonumber\\
&&\hspace{-90mm}{\ln(a_4^2e^{\phi})}''-6a_1^2a_2^2a_3^2\:e^{2\phi}=0,
\end{eqnarray}
which admit the following general solution:
\begin{eqnarray}
&&\hspace{-70mm}a_1^2e^{\phi}=\frac{L_1(\frac{p_4}{3})^{\frac{2}{3}}\,e^{p_1\tau}}{\sinh(p_4\tau)^{\frac{2}{3}}},\nonumber\\
&&\hspace{-70mm}a_2^2e^{\phi}=\frac{L_2(\frac{p_4}{3})^{\frac{2}{3}}\,e^{p_2\tau}}{\sinh(p_4\tau)^{\frac{2}{3}}},\nonumber\\
&&\hspace{-70mm}a_3^2e^{\phi}=\frac{(\frac{p_4}{3})^{\frac{2}{3}}\,e^{\left(N-p_1-p_2\right)\tau}}{(L_1L_2)\sinh(p_4\tau)^{\frac{2}{3}}},\nonumber\\
&&\hspace{-70mm}a_4^2e^{\phi}=\frac{(\frac{p_4}{3})^{\frac{2}{3}}\,
e^{\frac{N}{3}\tau}}{\sinh(p_4\tau)^{\frac{2}{3}}},
\end{eqnarray}
where $L_1$, $L_2$, $p_1$, $p_2$ and $p_4$ are constants. From the
initial value equation (19) we have the following restriction on
$p_i$:
\begin{eqnarray}
&&\hspace{-50mm}p_1^2+p_2^2+N^2+p_1p_2-N(p_1+p_2)-\frac{8}{3}p_4^2=0.
\end{eqnarray}
Note that the results for this algebra are same as for Lie algebra
$A_{4,5}^{(a,1)}$ for $a=1$.

\subsection{\bf Lie algebras $A_{4,6}^{(a,b)}$}
We obtain the potentials $V_i$ as follows:
\begin{eqnarray}
&&\hspace{-20mm}V_1=-\frac{a(a+2b)}{a_4^2},\:\:V_2=\frac{a_2^2}{2a_3^2a_4^2}-\frac{a_3^2}{2a_2^2a_4^2}-\frac{b(a+2b)}{a_4^2},\:\nonumber\\
&&\hspace{-20mm}V_3=-\frac{a_2^2}{2a_3^2a_4^2}+\frac{a_3^2}{2a_2^2a_4^2}-\frac{b(a+2b)}{a_4^2},\:\:
V_4=-\frac{a_2^2}{2a_3^2a_4^2}-\frac{a_3^2}{2a_2^2a_4^2}-\frac{(a^2+2b^2-1)}{a_4^2}.
\end{eqnarray}
The constraint equations (10) impose in this case the restriction
\begin{eqnarray}
&&\hspace{-120mm}a_1^{a}(a_2a_3)^{b}=Ba_4^{a+2b}.
\end{eqnarray}
where $B$ is a constant. Equations (18) have the following forms:
\begin{eqnarray}
&&\hspace{-70mm}{\ln(a_1^2 e^{\phi})}''-2\left(a^2+2ab\right)a_1^2a_2^2a_3^2e^{2\phi}=0,\nonumber\\
&&\hspace{-70mm}{\ln(a_2^2e^{\phi})}''-a_1^2\left(a_3^4-a_2^4+2b(2b^2+a)a_2^2a_3^2\right)e^{2\phi}=0,\nonumber\\
&&\hspace{-70mm}{\ln(a_3^2e^{\phi})}''-a_1^2\left(a_2^4-a_3^4+2b(2b^2+a)a_2^2a_3^2\right)e^{2\phi}=0,\nonumber\\
&&\hspace{-70mm}{\ln(a_4^2e^{\phi})}''-a_1^2\left(a_2^4+a_3^4+2(2b^2+a^2-1)a_2^2a_3^2\right)e^{2\phi}=0,
\end{eqnarray}
which for $a\neq 2b$ admit the following general solution:
\begin{eqnarray}
&&\hspace{-70mm}a_1^2e^{\phi}=\frac{L_1p_1\,e^{p_1\tau}}{\sinh(p_1\tau)^{\frac{2a}{a+2b}}},\nonumber\\
&&\hspace{-70mm}a_2^2e^{\phi}=a_3^2e^{\phi}=\frac{\sqrt{p_1}\,e^{(\frac{N-p_1)}{2})\tau}}{\sqrt{L_1}(a+2b)\sinh(p_1\tau)^{\frac{2b}{a+2b}}},\nonumber\\
&&\hspace{-70mm}a_4^2e^{\phi}=\left(\frac{L_1^{a-b}}{B^2}\frac{p_1^{a+b}\,e^{\left((a-b)p_1+bN\right)\tau}}{\sinh(p_1\tau)^{2(\frac{a^2+2b^2}{a+2b})}}\right)^{\frac{1}{a+2b}},
\end{eqnarray}
where $L_1$ and $p_1$ are constants. From the initial value equation
(19) we have the following restriction on $p_1$:
\begin{eqnarray}
&&\hspace{-10mm}(13a^2+36b^2+20ab)p_1^2-(3a^2+4b^2+8ab)N^2+6a(a+2b)Np_1=0.
\end{eqnarray}
But for the case $a=2b$, its solution is similar to $VI_0\oplus R$
model solution.

\subsection{\bf Lie algebras $A_{4,7}$}
We obtain the potentials $V_i$ as follows:
\begin{eqnarray}
&&\hspace{-70mm}V_1=\frac{a_1^2}{2a_2^2a_3^2}-\frac{8}{a_4^2},\:\:V_2=-\frac{a_1^2}{2a_2^2a_3^2}+\frac{a_2^2}{2a_3^2a_4^2}-\frac{4}{a_4^2},\nonumber\\
&&\hspace{-70mm}V_3=-\frac{a_1^2}{2a_2^2a_3^2}-\frac{a_2^2}{2a_3^2a_4^2}-\frac{4}{a_4^2},\:\:V_4=-\frac{a_2^2}{2a_3^2a_4^2}-\frac{6}{a_4^2}.
\end{eqnarray}
In this case the constraints imposed by (10) and (11) are
\begin{eqnarray}
&&\hspace{-110mm}a_1^2a_2a_3=B a_4^4,\nonumber\\
&&\hspace{-110mm}a_2=0,
\end{eqnarray}
where $B$ is constant; so that, within the present context, all
solutions are singular.

\subsection{\bf Lie algebras $A_{4,8}$}
We obtain the potentials $V_i$ as follows:
\begin{eqnarray}
&&\hspace{-70mm}V_1=\frac{a_1^2}{2a_2^2a_3^2},\:\:V_2=-\frac{a_1^2}{2a_2^2a_3^2},\:
V_3=-\frac{a_1^2}{2a_2^2a_3^2},\:\:V_4=-\frac{2}{a_4^2}.
\end{eqnarray}
The constraint equations (10) impose in this case the restriction
\begin{eqnarray}
&&\hspace{-120mm}a_2=Ba_3,
\end{eqnarray}
where $B$ is constant. In this case equations (18) have the
following forms:
\begin{eqnarray}
&&\hspace{-70mm}{\ln(a_1^2 e^{\phi})}''+a_1^4a_4^2e^{2\phi}=0,\nonumber\\
&&\hspace{-70mm}{\ln(a_2^2e^{\phi})}''-a_1^4a_4^2e^{2\phi}=0,\nonumber\\
&&\hspace{-70mm}{\ln(a_3^2e^{\phi})}''-a_1^4a_4^2e^{2\phi}=0,\nonumber\\
&&\hspace{-70mm}{\ln(a_4^2 e^{\phi})}''-4a_1^2a_2^2a_3^2e^{2\phi}=0,
\end{eqnarray}
which admit the following general solution:
\begin{eqnarray}
&&\hspace{-70mm}a_1^2e^{\phi}=\frac{3p_1\,e^{p_1\tau}}{2(L_1B)^2}\sinh(p_1\tau)^2,\nonumber\\
&&\hspace{-70mm}a_2^2e^{\phi}=\frac{L_1B^2\sqrt{p_1}\,e^{(\frac{N-p_1}{2})\tau}}{\sinh(p_1\tau)^2},\nonumber\\
&&\hspace{-70mm}a_3^2e^{\phi}=\frac{L_1\sqrt{p_1}\,e^{(\frac{N-p_1}{2})\tau}}{\sinh(p_1\tau)^2},\nonumber\\
&&\hspace{-70mm}a_4^2e^{\phi}=\frac{8(L_1B)^4\,e^{(N-2p_1)\tau}}{9\sinh(p_1\tau)^6},
\end{eqnarray}
where $L_1$ and $p_1$ are constants. From the initial value equation
(19) we have the following restriction on $p_1$:
\begin{eqnarray}
&&\hspace{-70mm}3N^2-29p_1^2+2Np_1=0.
\end{eqnarray}

\subsection{\bf Lie algebras $A_{4,9}^{b} \,\, (b\neq \pm 1,-2)$}
We obtain the potentials $V_i$ as follows:
\begin{eqnarray}
&&\hspace{-70mm}V_1=\frac{a_1^2}{2a_2^2a_3^2}-\frac{2(b+1)^2}{a_4^2},\:\:V_2=-\frac{a_1^2}{2a_2^2a_3^2}-\frac{2(b+1)}{a_4^2},\:\nonumber\\
&&\hspace{-70mm}V_3=-\frac{a_1^2}{2a_2^2a_3^2}-\frac{2b(b+1)}{a_4^2},\:\:V_4=-\frac{2(b^2+b+1)}{a_4^2}.
\end{eqnarray}
To facilitate calculations we will assume for the moment that we
also have $b\neq\pm1,-2$ and present the $A_{4,9}^{b=\pm 1,2}$
models separately in the next subsection. In this case the
constraint equations (10) impose the following restriction:
\begin{eqnarray}
&&\hspace{-120mm}a_1^{b+1}a_2a_3^{b}=Ba_4^{2(b+1)},
\end{eqnarray}
where $B$ is constant. Equations (18) have the following forms:
\begin{eqnarray}
&&\hspace{-70mm}{\ln(a_1^2 e^{\phi})}''+\left(a_1^4a_4^2-4(b+1)^2a_1^2a_2^2a_3^2\right)e^{2\phi}=0,\nonumber\\
&&\hspace{-70mm}{\ln(a_2^2e^{\phi})}''-\left(a_1^4a_4^2+4(b+1)a_1^2a_2^2a_3^2\right)e^{2\phi}=0,\nonumber\\
&&\hspace{-70mm}{\ln(a_3^2e^{\phi})}''-\left(a_1^4a_4^2+4b(b+1)a_1^2a_2^2a_3^2\right)e^{2\phi}=0,\nonumber\\
&&\hspace{-70mm}{\ln(a_4^2e^{\phi})}''-4\left(b^2+b+1\right)a_1^2a_2^2a_3^2e^{2\phi}=0.
\end{eqnarray}
It follows that for $b\neq -2$ and $b\neq\pm 1$ the most general
solution for these equations which satisfies the above constraint
is:
\begin{eqnarray}
&&\hspace{-70mm}a_1^2e^{\phi}=\frac{L_1\,e^{p_1\tau}}{\sinh(p_1\tau)^{\frac{4b^2+10b+4}{7b^2+13b+7}}},\nonumber\\
&&\hspace{-70mm}a_2^2e^{\phi}=\frac{L_2\,e^{\left(-\frac{b+2}{b-1}N+\frac{4b+5}{b-1}p_1\right)\tau}}{\sinh(p_1\tau)^{\frac{2b^2+8b+8}{7b^2+13b+7}}},\nonumber\\
&&\hspace{-70mm}a_3^2e^{\phi}=\frac{L_3\,e^{\left(-\frac{5b^2+14b+8}{(b+2)(b-1)}p_1+\frac{2b+1}{b-1}N\right)\tau}}{\sinh(p_1\tau)^{\frac{8b^2+8b+2}{7b^2+13b+7}}},\nonumber\\
&&\hspace{-70mm}a_4^2e^{\phi}=\frac{L_4\,e^{(N-2p_1)\tau}}{\sinh(p_1\tau)^{\frac{6(b^2+b+1)}{7b^2+13b+7}}},
\end{eqnarray}
where $L_1$ and $p_1$ are constants and the coefficient $L_2$, $L_3$
and $L_4$ are obtained as follows:
\begin{eqnarray}
&&\hspace{-70mm}L_2=\left(\frac{(3/2)^{b}L_1^{2b+3}}{{B}^2p_1^{2(b+1)}}\frac{(7b^2+13b+7)^{b+2}}{\left(2(b^2+b+1)\right)^{2(b+1)}}\right)^{\frac{1}{b-1}},\nonumber\\
&&\hspace{-70mm}L_3=\left(\frac{(2/3){B}^2p_1^{4b+2}}{L_1^{3b+2}}\frac{\left(2(b^2+b+1)\right)^{2(b+1)}}{(7b^2+13b+7)^{2b+1}}\right)^{\frac{1}{b-1}},\nonumber\\
&&\hspace{-70mm}L_4=\frac{2(b^2+b+1)}{(7b^2+13b+7)}\frac{p_1^2}{L_1^2}.
\end{eqnarray}
From the initial value equation (19) we have the following
restriction on $p_1$:
\begin{eqnarray}
&&\hspace{-25.5mm}\left(\frac{3(7b^2+13b+7)}{(b-1)^2}-\frac{4(5b^2+8b+5)}{(7b^2+13b+7)}\right)p_1^2+
\left(\frac{3(b^2+b+1)}{(b-1)^2}\right)N^2\nonumber\\
&&\hspace{-17.5mm}-\frac{2(7b^2+13b+7)}{(b-1)^2}Np_1=0.
\end{eqnarray}

\subsubsection{\bf Lie algebras $A_{4,9}^{b}\:\:(b=\pm 1,-2)$}
\textit{i)} For $b=1$ the constraint equation (114) may be written
as
\begin{eqnarray}
&&\hspace{-120mm}a_1^{2}a_2a_3=Ba_4^{4},
\end{eqnarray}
so that Eqs.(115) reduce to
\begin{eqnarray}
&&\hspace{-70mm}{\ln(a_1^2 e^{\phi})}''+\left(a_1^4a_4^2-16a_1^2a_2^2a_3^2\right)e^{2\phi}=0,\nonumber\\
&&\hspace{-70mm}{\ln(a_2^2e^{\phi})}''-\left(a_1^4a_4^2+8a_1^2a_2^2a_3^2\right)e^{2\phi}=0,\nonumber\\
&&\hspace{-70mm}{\ln(a_3^2e^{\phi})}''-\left(a_1^4a_4^2+8a_1^2a_2^2a_3^2\right)e^{2\phi}=0,\nonumber\\
&&\hspace{-70mm}{\ln(a_4^2e^{\phi})}''-12a_1^2a_2^2a_3^2e^{2\phi}=0.
\end{eqnarray}
The general solution to these equations are
\begin{eqnarray}
&&\hspace{-70mm}a_1^2e^{\phi}=\frac{L_1\,e^{\frac{N}{3}\tau}}{\sinh^{\frac{2}{3}}(p_1\tau)},\nonumber\\
&&\hspace{-70mm}a_2^2e^{\phi}=\frac{L_2\,e^{p_2\tau}}{\sinh^{\frac{2}{3}}(p_1\tau)},\nonumber\\
&&\hspace{-70mm}a_3^2e^{\phi}=\frac{L_3\,e^{(\frac{2N}{3}-p_2)\tau}}{\sinh^{\frac{2}{3}}(p_1\tau)},\nonumber\\
&&\hspace{-70mm}a_4^2e^{\phi}=\frac{L_4\,e^{\frac{N}{3}\tau}}{\sinh^{\frac{2}{3}}(p_1\tau)},
\end{eqnarray}
where $L_1$, $p_1$, $p_2$ are constants and the coefficient $L_2$,
$L_3$ and $L_4$ are obtained as follows:
\begin{eqnarray}
&&\hspace{-100mm}L_1=(32{B}^2)^{\frac{1}{9}}(\frac{p_1}{3})^{\frac{2}{3}},\nonumber\\
&&\hspace{-100mm}L_3=\frac{1}{36L_2}(\frac{16}{B^2})^{\frac{1}{9}}(3p_1^2)^{\frac{2}{3}},\nonumber\\
&&\hspace{-100mm}L_4=\frac{(p_1/3)^{\frac{2}{3}}}{(2{B}^4)^{\frac{1}{9}}}.
\end{eqnarray}
From the initial value equation (19) we have the following
restriction on $p_2$:
\begin{eqnarray}
&&\hspace{-90mm}54p_2^2-58N^2-72Np_2-27N=0.
\end{eqnarray}
\textit{ii)} For $b=-1$ the constraint equation (114) may be written
as
\begin{eqnarray}
&&\hspace{-120mm}a_1a_2=a_3,
\end{eqnarray}
and equations (115) get the following forms:
\begin{eqnarray}
&&\hspace{-70mm}{\ln(a_1^2 e^{\phi})}''+a_1^4a_4^2e^{2\phi}=0,\nonumber\\
&&\hspace{-70mm}{\ln(a_2^2e^{\phi})}''-a_1^4a_4^2e^{2\phi}=0,\nonumber\\
&&\hspace{-70mm}{\ln(a_3^2e^{\phi})}''-a_1^4a_4^2e^{2\phi}=0,\nonumber\\
&&\hspace{-70mm}{\ln(a_4^2 e^{\phi})}''-4a_1^2a_2^2a_3^2e^{2\phi}=0,
\end{eqnarray}
which admit the following general solution:
\begin{eqnarray}
&&\hspace{-70mm}a_1^2e^{\phi}=L_1\,e^{p_1\tau}\sinh^2(p_1\tau),\nonumber\\
&&\hspace{-70mm}a_2^2e^{\phi}=\sqrt{\frac{3}{2L_1}}\frac{p_1\,e^{\left(\frac{N-p_1}{2}\right)\tau}}{\sinh^2(p_1\tau)},\nonumber\\
&&\hspace{-70mm}a_3^2e^{\phi}=\sqrt{\frac{3}{2L_1}}\frac{p_1\,e^{\left(\frac{N-p_1}{2}\right)\tau}}{\sinh^2(p_1\tau)},\nonumber\\
&&\hspace{-70mm}a_4^2e^{\phi}=\frac{2p_1^2\,e^{\left(N-2p_1\right)\tau}}{L_1^2\sinh^2(p_1\tau)},
\end{eqnarray}
where $L_1$ and $p_1$ are constants. From the initial value equation
(19) we have the following restriction on $p_1$:
\begin{eqnarray}
&&\hspace{-90mm}3N^2-29p_1^2-2Np_1=0.
\end{eqnarray}
\textit{iii)} For case $b=-2$ the equations (114, 115) can be reduce
to
\begin{eqnarray}
&&\hspace{-120mm}a_2a_4^{2}=Ba_1a_3^2,
\end{eqnarray}
and
\begin{eqnarray}
&&\hspace{-70mm}{\ln(a_1^2 e^{\phi})}''+\left(a_1^4a_4^2-4a_1^2a_2^2a_3^2\right)e^{2\phi}=0,\nonumber\\
&&\hspace{-70mm}{\ln(a_2^2e^{\phi})}''-\left(a_1^4a_4^2-4a_1^2a_2^2a_3^2\right)e^{2\phi}=0,\nonumber\\
&&\hspace{-70mm}{\ln(a_3^2e^{\phi})}''-\left(a_1^4a_4^2+8a_1^2a_2^2a_3^2\right)e^{2\phi}=0,\nonumber\\
&&\hspace{-70mm}{\ln(a_4^2e^{\phi})}''-12a_1^2a_2^2a_3^2e^{2\phi}=0,
\end{eqnarray}
which admit the following solutions:
\begin{eqnarray}
&&\hspace{-70mm}a_1^2e^{\phi}=L_1\,e^{p_1\tau},\nonumber\\
&&\hspace{-70mm}a_2^2e^{\phi}=(\frac{L_1{B}^2}{16})^{\frac{1}{3}}e^{p_1\tau},\nonumber\\
&&\hspace{-70mm}a_3^2e^{\phi}=(\frac{2}{27A^4{B}^2})^{\frac{1}{3}}\frac{p_1^2\,e^{(N-2p_1)\tau}}{\sinh^2(p_1\tau)},\nonumber\\
&&\hspace{-70mm}a_4^2e^{\phi}=\frac{2p_1^2\,e^{(N-2P_1)\tau}}{3L_1^2\sinh^2(p_1\tau)},
\end{eqnarray}
where $L_1$ and $p_1$ are constants. From the initial value equation
(19) we have the following restriction on $p_1$:
\begin{eqnarray}
&&\hspace{-90mm}N^2-p_1^2+2Np_1=0.
\end{eqnarray}

\subsection{\bf Lie algebras $A_{4,9}^{0}$}
We obtain the potentials $V_i$ as follows:
\begin{eqnarray}
&&\hspace{-40mm}V_1=\frac{a_1^2}{2a_2^2a_3^2}-\frac{2}{a_4^2},\:\:V_2=-\frac{a_1^2}{2a_2^2a_3^2}-\frac{2}{a_4^2},\:\:
V_3=-\frac{a_1^2}{2a_2^2a_3^2},\:\:V_4=-\frac{2}{a_4^2}.
\end{eqnarray}
In this case the constraint equations (10) impose the following
restriction:
\begin{eqnarray}
&&\hspace{-120mm}a_1a_2=Ba_4^{2},
\end{eqnarray}
where $B$ is constant. Equations (18) have the following forms:
\begin{eqnarray}
&&\hspace{-70mm}{\ln(a_1^2 e^{\phi})}''+\left(a_1^4a_4^2-4a_1^2a_2^2a_3^2\right)e^{2\phi}=0,\nonumber\\
&&\hspace{-70mm}{\ln(a_2^2e^{\phi})}''-\left(a_1^4a_4^2+4a_1^2a_2^2a_3^2\right)e^{2\phi}=0,\nonumber\\
&&\hspace{-70mm}{\ln(a_3^2e^{\phi})}''-a_1^4a_4^2e^{2\phi}=0,\nonumber\\
&&\hspace{-70mm}{\ln(a_4^2 e^{\phi})}''-4a_1^2a_2^2a_3^2e^{2\phi}=0,
\end{eqnarray}
which admit the following general solutions:
\begin{eqnarray}
&&\hspace{-90mm}a_1^2e^{\phi}=\frac{L_1\,e^{p_1\tau}}{\sinh^{\frac{4}{7}}(p_1\tau)},\nonumber\\
&&\hspace{-90mm}a_2^2e^{\phi}=\frac{4p_1^4{B}^{2}\,e^{(2N-5p_1)\tau}}{49L_1^5\sinh^{\frac{8}{7}}(p_1\tau)},\nonumber\\
&&\hspace{-90mm}a_3^2e^{\phi}=\frac{21L_1^4e^{(4p_1-N)\tau}}{8B^2p_1^2\sinh^{\frac{2}{7}}(p_1\tau)},\nonumber\\
&&\hspace{-90mm}a_4^2e^{\phi}=\frac{2p_1^2\,e^{(N-2p_1)\tau}}{7L_1^2\sinh^{\frac{6}{7}}(p_1\tau)},
\end{eqnarray}
where $L_1$ and $p_1$ are constants. From the initial value equation
(19) we have the following restriction on $p_1$:
\begin{eqnarray}
&&\hspace{-100mm}21N^2+127p_1^2 -98Np_1=0.
\end{eqnarray}

\subsection{\bf Lie algebras $A_{4,10}$}
We obtain the potentials $V_i$ as follows:
\begin{eqnarray}
&&\hspace{-50mm}V_1=-\frac{a_1^2}{2a_2^2a_3^2},\:\:V_2=\frac{a_1^2}{2a_2^2a_3^2}-\frac{a_2^2}{2a_3^2a_4^2}+\frac{a_3^2}{2a_2^2a_4^2},\nonumber\\
&&\hspace{-50mm}V_3=\frac{a_1^2}{2a_2^2a_3^2}+\frac{a_2^2}{2a_3^2a_4^2}-\frac{a_3^2}{2a_2^2a_4^2},\:\:V_4=\frac{a_2^2}{2a_3^2a_4^2}+\frac{a_3^2}{2a_2^2a_4^2}-\frac{1}{a_4^2}.
\end{eqnarray}
such that there are no constraint equation. Equations (18) have
the following forms:
\begin{eqnarray}
&&\hspace{-70mm}{\ln(a_1^2 e^{\phi})}''-a_1^4a_4^2e^{2\phi}=0,\nonumber\\
&&\hspace{-70mm}{\ln(a_2^2e^{\phi})}''+\left(a_1^2a_3^4-a_1^2a_2^4+a_4^2a_1^4\right)e^{2\phi}=0,\nonumber\\
&&\hspace{-70mm}{\ln(a_3^2e^{\phi})}''+\left(a_1^2a_2^4-a_1^2a_3^4+a_4^2a_1^4\right)e^{2\phi}=0,\nonumber\\
&&\hspace{-70mm}{\ln(a_4^2e^{\phi})}''+\left(a_1^2a_3^4-2a_1^2a_2^2a_3^2+a_1^2a_2^4\right)e^{2\phi}=0,
\end{eqnarray}
which admit the following general solution:
\begin{eqnarray}
&&\hspace{-70mm}a_1^2e^{\phi}=\frac{p_1}{\sqrt{L_4}\sinh(p_1\tau)},\nonumber\\
&&\hspace{-70mm}a_2^2e^{\phi}=a_3^2e^{\phi}=L_2e^{p_2\tau}\sinh(p_1\tau),\nonumber\\
&&\hspace{-70mm}a_4^2e^{\phi}=L_4e^{N\tau},
\end{eqnarray}
where $L_2$, $L_4$, $p_1$ and $p_2$ are constants. From the initial
value equation (19) we have the following restriction on $p_1$ and
$p_2$:
\begin{eqnarray}
&&\hspace{-70mm}p_1^2-p_2^2+N^2=0.
\end{eqnarray}

\subsection{\bf Lie algebras $A_{4,11}^{a}$}
We obtain the potentials $V_i$ as follows:
\begin{eqnarray}
&&\hspace{-30mm}V_1=\frac{a_1^2}{2a_2^2a_3^2}-\frac{8b^2}{a_4^2},\:\:
V_2=-\frac{a_1^2}{2a_2^2a_3^2}+\frac{a_2^2}{2a_3^2a_4^2}-\frac{a_3^2}{2a_2^2a_4^2}-\frac{4b^2}{a_4^2},\nonumber\\
&&\hspace{-30mm}V_3=-\frac{a_1^2}{2a_2^2a_3^2}-\frac{a_2^2}{2a_3^2a_4^2}+\frac{a_3^2}{2a_2^2a_4^2}-\frac{4b^2}{a_4^2},\:\:
V_4=-\frac{a_2^2}{2a_3^2a_4^2}-\frac{a_3^2}{2a_2^2a_4^2}-\frac{6b^2-1}{a_4^2}.
\end{eqnarray}
The constraint equations (10) impose in this case the restriction
\begin{eqnarray}
&&\hspace{-120mm}a_1^2a_2a_3=Ba_4^4,
\end{eqnarray}
where $B$ is a constant. Equations (18) have the following forms:
\begin{eqnarray}
&&\hspace{-70mm}{\ln(a_1^2 e^{\phi})}''+\left(a_1^4a_4^2-16b^2a_1^2a_2^2a_3^2\right)e^{2\phi}=0,\nonumber\\
&&\hspace{-70mm}{\ln(a_2^2e^{\phi})}''-\left(a_3^4-a_2^4+a_1^2a_4^2+8b^2a_2^2a_3^2\right)a_1^2e^{2\phi}=0,\nonumber\\
&&\hspace{-70mm}{\ln(a_3^2e^{\phi})}''-\left(a_2^4-a_3^4+a_1^2a_4^2+8b^2a_2^2a_3^2\right)a_1^2e^{2\phi}=0,\nonumber\\
&&\hspace{-70mm}{\ln(a_4^2e^{\phi})}''-\left(a_2^4+a_3^4+(12b^2-2)a_2^2a_3^2\right)a_1^2e^{2\phi}=0.
\end{eqnarray}
These equations admit the following general solutions:
\begin{eqnarray}
&&\hspace{-70mm}a_1^2e^{\phi}=\frac{(\frac{C}{b})^{\frac{2}{3}}\,e^{\frac{N\tau}{3}}}{2\sinh^{\frac{2}{3}}(p_1\tau)},\nonumber\\
&&\hspace{-70mm}a_2^2e^{\phi}=a_3^2e^{\phi}=\frac{(\frac{1}{\sqrt{C}b})^{\frac{2}{3}}\,e^{\frac{N\tau}{3}}}{3\sinh^{\frac{2}{3}}(p_1\tau)},\nonumber\\
&&\hspace{-70mm}a_4^2e^{\phi}=\frac{(b^2\sqrt{C})^{\frac{1}{3}}\,e^{\frac{N\tau}{3}}}{\sqrt{6B}\sinh^{\frac{2}{3}}(p_1\tau)},
\end{eqnarray}
where $p_1$ is a constant and $C=\frac{4}{3}(2Ba^{\frac{4}{3}})^{\frac{1}{3}}$.\\
From the initial value equation (19) we have the following
restriction on $p_1$:
\begin{eqnarray}
&&\hspace{-70mm}N=\pm 2p_1.
\end{eqnarray}

\subsection{\bf Lie algebras $A_{4,12}$}
We obtain the potentials $V_i$ as follows:
\begin{eqnarray}
&&\hspace{-70mm}V_1=\frac{a_1^2}{2a_2^2a_4^2}-\frac{a_2^2}{2a_1^2a_4^2}-\frac{2}{a_3^2},\:\:V_2=-\frac{a_1^2}{2a_2^2a_4^2}+\frac{a_2^2}{2a_1^2a_4^2}-\frac{2}{a_3^2},\:\nonumber\\
&&\hspace{-70mm}V_3=-\frac{2}{a_3^2},\:\:V_4=-\frac{a_1^2}{2a_2^2a_4^2}-\frac{a_2^2}{2a_1^2a_4^2}+\frac{1}{a_4^2}.
\end{eqnarray}
In this case the constraint equations (10) impose the following
restriction:
\begin{eqnarray}
&&\hspace{-120mm}a_1a_2=Ba_3^2,
\end{eqnarray}
where $B$ is a constant. Equations (18) have the following forms:
\begin{eqnarray}
&&\hspace{-70mm}{\ln(a_1^2 e^{\phi})}''+\left(a_1^4a_3^2-a_2^4a_3^2-4a_1^2a_2^2a_4^2\right)e^{2\phi}=0,\nonumber\\
&&\hspace{-70mm}{\ln(a_2^2e^{\phi})}''-\left(a_1^4a_3^2-a_2^4a_3^2+4a_1^2a_2^2a_4^2\right)e^{2\phi}=0,\nonumber\\
&&\hspace{-70mm}{\ln(a_3^2e^{\phi})}''-4a_1^2a_2^2a_4^2\:e^{2\phi}=0,\nonumber\\
&&\hspace{-70mm}{\ln(a_4^2e^{\phi})}''-\left(a_1^4a_3^2+a_2^4a_3^2-2a_1^2a_2^2a_3^2\right)e^{2\phi}=0,
\end{eqnarray}
which admit the following general solution:
\begin{eqnarray}
&&\hspace{-70mm}a_1^2e^{\phi}=a_2^2e^{\phi}=\frac{p_1\,e^{(\frac{N-p_4}{2})\tau}}{2\sqrt{L_4}\sinh(p_1\tau)},\nonumber\\
&&\hspace{-70mm}a_3^2e^{\phi}=\frac{p_1\,e^{(\frac{N-p_4}{2})\tau}}{2B\sqrt{L_4}\sinh(p_1\tau)},\nonumber\\
&&\hspace{-70mm}a_4^2e^{\phi}=L_4e^{p_4\tau},
\end{eqnarray}
where $L_4$, $p_1$ and $p_4$ are constants. From the initial value
equation (19) we have the following restriction on $p_1$ and $p_4$:
\begin{eqnarray}
&&\hspace{-70mm}3N^2-12p_1^2+3p_4^2-2Np_4=0.
\end{eqnarray}

\section{\bf Example $VII_0\oplus R$ as a 4+1 dimensional cosmology coupled to matter with negative pressure (accelerating universe)}

In this section we apply one of the exact solutions of the previous
section (e.g $VII_0\oplus R$) as a physical cosmological model. In
Ref. \cite{Ba}, it has been shown that the equations (2-4) are equivalent to
the following Einstein equation
\begin{eqnarray}
&&\hspace{-70mm}\tilde{R}_{\mu\nu}-\frac{1}{2}\tilde{R}\tilde{g_{\mu\nu}}=\kappa^2(T^{(\phi)}_{\mu\nu}+T^{(H)}_{\mu\nu})
\end{eqnarray}
with the metric tensor
\begin{eqnarray}
&&\hspace{-100mm}\tilde{g_{\mu\nu}}=e^{\phi}g_{\mu\nu},
\end{eqnarray}
and the following energy-momentum tensor for $\phi$ and $H$
\begin{eqnarray}
&&\hspace{-70mm}\kappa^2T^{(\phi)}_{\mu\nu}=\frac{1}{2}\left(\partial_{\mu}\phi\partial_{\nu}\phi-\frac{1}{2}(\partial\phi)^2\tilde{g_{\mu\nu}}-\Lambda
e^{-\phi}\tilde{g_{\mu\nu}}\right),\nonumber\\
&&\hspace{-70mm}\kappa^2T^{(H)}_{\mu\nu}=\frac{1}{4}\left(H_{\mu\kappa\lambda}H_{\nu}^{\kappa\lambda}-\frac{1}{6}H^2\tilde{g_{\mu\nu}}\right).
\end{eqnarray}

Now, we obtain the components of above tensors and Einstein equation for
the example $VII_0\oplus R$ of subsection 4. For this propose, we
must obtain the metric $g_{\mu\nu}$ as functions of coordinate time
$t$. From relations (13), (15) and (54) we obtain the following
relation for $\tau$ as a function of $t$:
\begin{eqnarray}
&&\hspace{-70mm}\tau=(\frac{1}{\gamma})\ln(\frac{\gamma
t}{\sqrt{L_1L_2L_3}}),
\end{eqnarray}
where $\gamma=\frac{(p_1+p_3+p_4-2N)}{2}$. After substituting (154) in
relation (6) and using results of table $1$ the metric tensor can be
obtain as follows:
\begin{eqnarray}
&&\hspace{-40mm}ds^2=-dt^2+a_1(t)^2\left((dx^1)^2+(dx^2)^2\right)+a_3(t)^2(dx^3)^2+a_4(t)^2(dx^4)^2.
\end{eqnarray}
Using above equation one can calculate the non-zero components of
energy-momentum tensor $T^{(\phi)}_{\mu\nu}$ for the $VII_0\oplus R$
model as follows\footnote{Note that for our examples of section 4,
$H_{\mu\nu\lambda}=0$ and consequently $T^{(H)}_{\mu\nu}=0$}:
\begin{eqnarray}
&&\hspace{-40mm}\kappa^2T_{00}=\frac{N^2}{2\gamma^2t^2}\left(1+\frac{1}{2}(\frac{\gamma
t}{\sqrt{L_1L_2L_3}})^{\frac{N}{\gamma}}\right),\nonumber\\
&&\hspace{-40mm}\kappa^2T_{11}=\kappa^2T_{22}=-\sqrt{L_1}\frac{N^2}{4\gamma^2t^2}\left(\frac{\gamma
t}{\sqrt{L_1L_2L_3}}\right)^{\frac{p_1}{2\gamma}},\nonumber\\
&&\hspace{-40mm}\kappa^2T_{33}=-L_2\frac{N^2}{4\gamma^2t^2}\left(\frac{\gamma
t}{\sqrt{L_1L_2L_3}}\right)^{\frac{p_3}{\gamma}},\nonumber\\
&&\hspace{-40mm}\kappa^2T_{44}=-L_3\frac{N^2}{4\gamma^2t^2}\left(\frac{\gamma
t}{\sqrt{L_1L_2L_3}}\right)^{\frac{p_4}{\gamma}}.
\end{eqnarray}
Now by comparing the components of $T_{\mu\nu}$ with the following
perfect fluid energy-momentum tensor:
\begin{eqnarray}
&&\hspace{-40mm}T=\left(
                    \begin{array}{ccccc}
                      -\rho(t) & 0 & 0 & 0 & 0 \\
                      0 & P(t) & 0 & 0 & 0 \\
                      0 & 0 & P(t)& 0 & 0 \\
                      0 & 0 & 0 & P(t) & 0 \\
                      0 & 0 & 0 & 0 & P(t) \\
                    \end{array}
                  \right)
\end{eqnarray}
and assuming $\left(\kappa=1,\frac{p_1}{2}=p_3=p_4,
\sqrt{L_1}=L_2=L_3\right)$ and using (55), one can obtain the
density and pressure quantities of the matter and scale factors as
follows:
\begin{eqnarray}
&&\hspace{-40mm}\rho(t)=-\frac{1}{2}\left(\frac{9}{t^2}+\frac{N^3}{6L_3^6}t\right),\nonumber\\
&&\hspace{-40mm}P(t)=-\frac{N^2}{4L_3^3}
\end{eqnarray}
\begin{eqnarray}
&&\hspace{-40mm}a_i(t)=\sqrt{\frac{3L_3^3}{Nt}},
\quad\quad\,(i=1,2,3,4).
\end{eqnarray}
We examine our model with a cosmological model of dark energy. Thus,
from equation (158) we insist that the our model corresponds to the
equation of state for dark energy as follows:
\begin{eqnarray}
&&\hspace{-40mm}P(t)=-\rho(t),
\end{eqnarray}
So, by taking condition (160) and assuming $N=L_3^2$ we can get
\begin{eqnarray}
&&\hspace{-40mm}H_0=\frac{1}{t}=\frac{1}{81}\left(-54+6\sqrt{6L_3^3+81}\right)^{\frac{1}{3}}
-\frac{L_3}{3\left(-54+6\sqrt{6L_3^3+81}\right)^{\frac{1}{3}}},
\end{eqnarray}
where $H_{0}$ is the Hubble parameter at epoch that the universe
accelerates in the context of cosmological constant regime. We can
fix the parameter $L_{3}$ in our model so that leads to
accelerating universe (see for example \cite{Darabi}). Also, we can
determine another cosmological parameter so-called the deceleration
parameters which has the following negative values
\begin{eqnarray}
&&\hspace{-40mm}q_i(t)=-\frac{\ddot{a}_i(t)a_i(t)}{\dot{a}_i(t)^2}=-3.
\end{eqnarray}
From the above equation it can be seen we have $q<0$. Thus, our
model predicts an accelerating  universe.

\section{\bf Conclusion}

As a consequences let us examine the effect of Abelian target space
duality on the above backgrounds. Consider the following sigma model
action:
\begin{eqnarray}
&&\hspace{-50mm}S=\int_{\Sigma}d^2\xi\left(\em{g}^{\alpha\beta}\partial_{\alpha}x^{\mu}\partial_{\beta}x^{\nu}G_{\mu\nu}+
\varepsilon^{\alpha\beta}B_{\mu\nu}\partial_{\alpha}x^{\mu}\partial_{\beta}x^{\nu}+R^{(2)}\phi\right)
\end{eqnarray}
such that the equations (2-4) are the zeros of its one loop beta
function anomaly. In the above equation $\em{g}$, $R^{(2)}$ are
metric and scalar curvature of the two dimensional world sheet
$\Sigma$ respectively. In order to obtain the dual background it is enough to
apply the Buscher's duality transformations \cite{Buscher} in the
isometry direction.

The isometry directions of the $4+1$ dimensional homogenous
anisotropic backgrounds are written in the table 1. Here we
consider an example for $A_{4,8}\cong H_4$ Lie algebra. One can
obtain all other dual backgrounds in the same way by using
Buschers's duality \cite{Buscher}. For the Lie algebra $A_{4,8}$
the metric on the $4+1$ dimensional space-time is obtained as
follows:
\begin{eqnarray}
&&\hspace{-10mm}ds^2=-dt^2+a_1^2dx_1^2+2a_1^2x_3dx_1dx_2+\left(a_1^2x_3^2+a_2^2e^{2x_4}\right)dx_2^2+a_3^2e^{-2x_4}dx_3^2+a_4^2dx_4^2
\end{eqnarray}
This metric has two isometry directions $x_1$ and $x_2$, such that
we have $[X_1,X_2]=0$. Now by use of the Buscher's duality
transformation \cite{Buscher} in the $x_1$ direction one can
obtain the following dual background:
\begin{eqnarray}
\hspace{-20mm}\left\{\begin{array}{cccccc}
\hspace{1mm}\tilde{G_{11}}=\frac{1}{a_1^2},\,\,\tilde{G_{12}}=\tilde{G_{13}}=\tilde{G_{14}}=0,\\& \\
\hspace{-4mm}\tilde{G_{22}}=a_2^2e^{2x_4},\:\:\tilde{G_{23}}=\tilde{G_{24}}=0,\\ & & \\
\hspace{-13mm}\tilde{G_{33}}=a_3^2e^{-2x_4},\:\:\tilde{G_{34}}=0,\\&&\\
\hspace{-40mm}\tilde{G_{44}}=a_4^2,
\end{array} \right.
\hspace{-5mm}\left\{\begin{array}{cccccc}
\hspace{1mm}\tilde{B_{12}}=x_3,\:\:\tilde{B_{13}}=\tilde{B_{14}}=0,\\& \\
\hspace{-19mm}\tilde{B_{23}}=\tilde{B_{24}}=0, & &
\hspace{-23mm}\tilde{B_{34}}=0,
\end{array} \right.
\end{eqnarray}
\begin{eqnarray}
\hspace{-30mm}H_{123}=1,\:\: \tilde{\phi}=\phi(t)-\frac{1}{2}\ln
a_1(t).
\end{eqnarray}
We see that in the dual background the antisymmetric tensor is
nonzero. In this way, in order to obtain the solutions of equations
of motions (2-4) for this backgrounds, it is enough to use of
duality transformations. In the similar way one can obtain other
dual backgrounds by considering isometry directions mentioned in the
table 1 and using Buscher's duality transformations in these
directions.\ \newline

\textbf{Acknowledgment}

The authors is grateful to F. Darabi and K. Atazadeh for useful
discussion and recommendation.

\end{document}